\documentclass[pra,aps,twocolumn,amsmath,amssymb,floatfix,showpacs]{revtex4}

\usepackage{graphicx}
\begin{document}

\title{Entanglement detection and quantum metrology by Raman photon diffraction imaging}

\author{Hongyi Yu} \author{Wang Yao}
\thanks{wangyao@hku.hk}

\affiliation{Department of Physics and Center of Theoretical and
  Computational Physics, The University of Hong Kong, Hong Kong,
  China}

\date{\today}

\begin{abstract}

  We show that far field diffraction image of spontaneously scattered
  Raman photons can be used for detection of spin entanglement and
  for metrology of fields gradients in cold atomic ensembles. For
  many-body states with small or maximum uncertainty in
  spin-excitation number, entanglement is simply witnessed by the
  presence of a sharp diffraction peak or dip. Gradient vector of
  external fields is measured by the displacement of a diffraction
  peak due to inhomogeneous spin precessions, which suggests a new
  possibility for precision measurement beyond the standard quantum
  limit without entanglement. Monitoring temporal decay of the
  diffraction peak can also realize non-demolition probe of
  temperature and collisional interactions in trapped cold atomic
  gases. The approach can be readily generalized to cold molecules,
  trapped ions, and solid state spin ensembles.

\end{abstract}


\maketitle

\section{Introduction}
Cold atomic ensembles offer an ideal platform for the study of quantum
many-body physics and for the implementation of quantum information
processing~\cite{Bloch_RMP}.  With entanglement speculated as a key
phenomenon in these occasions, efficient approach to detect
entanglement is crucial for understanding its profound
roles~\cite{Guhne_entanglementdetection}.  Spin of cold atoms is also
widely used for precision measurement of external fields. A topic of
current interest is quantum metrology which utilizes quantum
properties and particularly entanglement in the probe system to reach
measurement sensitivity beyond the standard quantum limit
(SQL)~\cite{Lloyd_metrology}.

To address these outstanding demands in the exploration of quantum
physics and quantum technology using cold atomic ensembles, the key is
efficient access to the right piece of information in the spin
subspace.  An ideal interface between spin and photon is offered by
the process of spontaneous Stokes
scattering~\cite{EIT,DLCZ,quantum_memory0,quantum_memory1,quantum_memory2,quantum_memory3,quantum_memory4,Pan_quantummemory,dephaseinOL1,dephaseinOL2}:
with a laser driving an ensemble of atoms in $\Lambda$-configuration,
a collective spin-excitation can be spontaneously converted into a
Stokes photon with phase and wavevector preserved. One may thus
anticipate that photon diffraction pattern can provide information on
collective spin properties. Earlier studies on the diffraction of
collectively emitted photons have focused on the super-radiance
phenomenon (i.e. induced directional coherent radiation) in very dense
atomic
ensembles~\cite{Eberly_superradiance,Haroche_superradiance,Carmichael_superradiance},
or in ensembles prepared with a single
excitation~\cite{Wstate,Scully_singlephotonsuperradiance1,Scully_singlephotonsuperradiance2,Eberly_singlephotonsuperradiance,Wstate2}.

In this paper, we show that the far field diffraction image of
spontaneously emitted Raman photons can be used for detection of spin
entanglement and for precision measurement of gradient vector of
external fields in cold atomic ensembles. We find the strength of a
sharp diffraction peak or dip measures spin pair-correlation sum and
detects entanglement through pair-correlation sum rules we derive from
optimal spin squeezing
inequalities~\cite{Sorensen_spinsqueezing,Korbicz_spinsqueezing,Toth_squeezing,
  Duan2011}. For many-body states with small or maximum uncertainty in
spin-excitation number, entanglement is simply witnessed by the
presence of the peak or dip. Inhomogeneous spin precessions in a field
gradient lead to displacement of the diffraction peak (dip), which can
serve as a principle for vector metrology of fields gradients and for
calibration of inhomogeneity in optical lattices. The gradiometer
sensitivity can reach $1/N$ by using a spin-coherent-state of $N$
unentangled atoms as the probe, which suggests a new possibility for
going beyond the SQL of $1/\sqrt{N}$ without
entanglement~\cite{fockstates1,fockstates2,HL1,HL2}.  Motional
dynamics leads to temporal decay of the diffraction peak which can be
used for non-demolition probe of temperature and collisional
interactions in trapped atomic gases.

Two remarkable features make this approach particularly suitable for
ensembles with large number of atoms. First, regardless of the
ensemble size, spin dephasing noise as a major error source only
results in decay of the peak (dip) strength in a timescale equal to
the dephasing time of a single spin. Second, the number of useful
photons from a single copy of many-body state can be as large as its
spin-excitation number for cold atomic ensembles which are typically
dilute (i.e. interatomic distance comparable to or larger than optical
wavelength). This approach complements existing optical methods for
probing many-body quantum
states~\cite{Altman,Eckert,Bruun,Vega,Corcovilos,Miyake,Weitenberg},
and is readily applicable in other systems including molecular
ensembles, trapped ions and solid state spin ensembles.

\begin{figure*}[tbp]
\includegraphics[width=12cm]{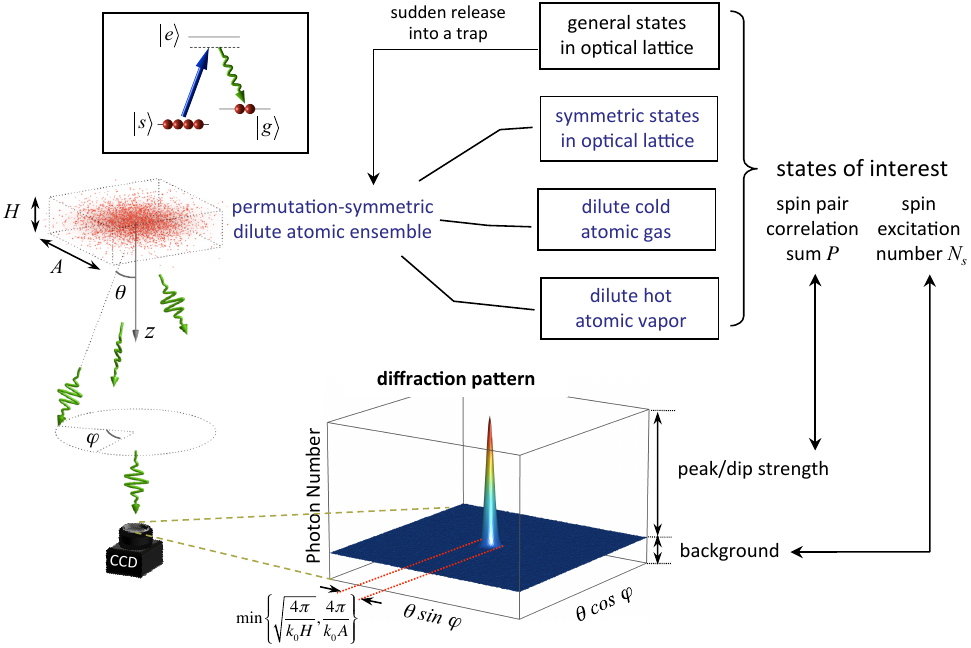}
\caption{\label{Fig1} Far field diffraction image of Stokes photons
  from permutation-symmetric dilute ensembles. The pair-correlation
  sum $P$ in the many-atom state of interest manifests as a sharp
  diffraction peak (for $P>0$) or dip (for $P<0$) along the forward
  direction, with strength $\propto |P|$ and width inversely
  proportional to the ensemble size. \label{fig1} }
\end{figure*}

The rest of the paper is organized as follows. In section II, we analyze the
the far field diffraction pattern of Raman photons and show how to extract the pair-correlation sum of atomic spins. In section III,  we derive pair-correlation sum rules for detecting entanglement. In section IV, we analyze the time evolution of the diffraction pattern from dilute ensembles. In section V, we discuss the use of
the diffraction pattern for precision measurement of field gradient and for non-demolition probe of atomic motion and temperature. Section VI is a
brief summary to the paper. More supplementary details on the derivations are grouped in the Appendices.

\section{Diffraction Pattern of Stokes Photons}

Consider an optically thin cold atomic ensemble with a $\Lambda$ level
configuration where two atomic ground states $|g\rangle$ and $|s
\rangle$ can be optically coupled to a common excited state $| e
\rangle$ (Fig.~1 inset). The ensemble is driven by a laser with Rabi
frequency $\Omega_L$, detuning $\Delta$ and wavevector $\mathbf k_0 =
k_0 \hat{\mathbf z} $. We assume atomic motion can be taken as frozen
in the duration of photon emission. With the laser coupling the $| s
\rangle$ to $|e \rangle$ transition, an atom can go from state
$|s\rangle$ to $|g\rangle$ by emitting a Stokes photon into the
vacuum. When $\Delta$ is much larger than $\Omega_L$ and the excited
state homogeneous line width $\Gamma_0$, $|e\rangle$ can be
adiabatically eliminated, leading to the effective light-atom coupling
in the electric-dipole and rotating wave approximation:  
\begin{eqnarray}
  \hat H & = & \sum_{\mathbf k} \hbar \omega_k \hat a_{\mathbf k}^\dag
  \hat a_{\mathbf k} + \sum_j E_z \hat \sigma_j^z \nonumber\\
  & & +\sum_{\mathbf k} g_{\mathbf k} \sum_j\mathrm e^{-i (\mathbf k -
    \mathbf k_0) \cdot \mathbf r_j}\hat \sigma_j^- \hat a_{\mathbf
    k}^\dag+\textrm{h.c.}.
  \label{light_atom_Hamiltonian}
\end{eqnarray}
Here $g_{\mathbf k} =
\frac{\Omega_L}{2\Delta}\sqrt{\frac{2\pi\omega_k}{V}}\hat{\mathbf
  e}_{\mathbf k} \cdot \boldsymbol {\mu}$, $\hat{\mathbf e}_{\mathbf
  k}$ and $\boldsymbol {\mu}$ being respectively the unit polarization
vector and the single atom dipole. $\hat \sigma_j^- \equiv | g
\rangle_j \langle s |$ and $\hat \sigma_j^z \equiv |s \rangle_j
\langle s | - |g \rangle_j \langle g |$. We assume anti-Stokes
scattering is either forbidden by the polarization selection rule or
suppressed by the much larger detuning when $E_z \gg \hbar\Delta$.

\begin{figure*}[tbp]
\includegraphics[width=11cm]{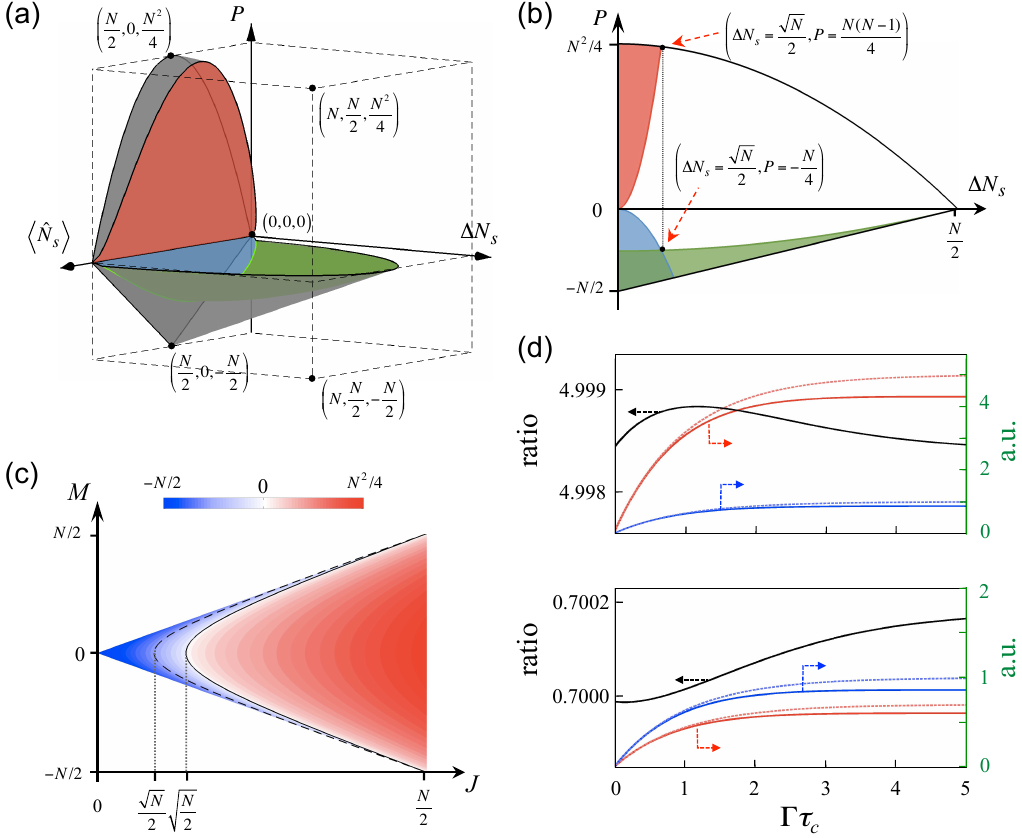}
\caption{\label{Fig2} (a) Phase diagram in the parameter space $(
  \langle \hat{N}_s \rangle, \Delta N_s, P )$. States in the
  surrounded region are all entangled ones.  (b) A slice of (a) taken
  for $\langle \hat{N}_s \rangle =N/2$.  The red, blue and green
  regions are entangled states violating inequalities
  (\ref{witness1}), (\ref{witness2}) and (\ref{witness3})
  respectively. The grey surfaces in (a) and the black curves in (b)
  are boundaries between physical and unphysical regions. Positive and
  negative sections of $P$ axis use different linear scale. (c)
  Strength of the diffraction peak (red) or dip (blue) for eigenstates
  of total spin $\hat{J}^2$ and $\hat{J}_z$. Inequalities
  (\ref{witness1}) and (\ref{witness2}) are violated in the peak and
  dip regions respectively.  States violating inequality
  (\ref{witness3}) form a subset of the dip region, to the left of the
  dashed curve.  (d) Upper (lower): peak (dip) to background ratio as
  a function of the collection interval $\tau_c$ for a
  half-spin-excitation state with $P=2.5N$ ($P=-0.35N$), shown as the
  black curve. The calculation is for $N=4000$ atoms of a 2D Gaussian
  distribution with FWHM $A=100~\mu$m. Peak or dip (background)
  strength is evaluated at $\theta=0$ ($\theta=\frac{2 \pi}{ k_0 A}
  $), shown by the blue (red) solid curve. Dashed curves are
  calculations with the multiple-light scattering and dipole-dipole
  interaction neglected. \label{fig2}}
\end{figure*}

Emission of a Stokes photon into mode $\mathbf{k} = (k, \theta,
\varphi) $ is accompanied by the annihilation of
a spin excitation by $ \hat J^-(\Delta \mathbf k) \equiv
\sum_{j=1}^N \mathrm e^{-i \Delta \mathbf k \cdot \mathbf r_j}\hat
\sigma_j^-$, $\Delta \mathbf k = \mathbf k- \mathbf k_0 $. The
angular distribution of the photon emission rate is given by $I
(\theta, \varphi, t) = I_{s} (\theta) I_{c} (\theta, \varphi, t)
\nonumber$. $I_s$ is the single atom dipole emission pattern, a slow
varying function of $\theta$. $I_c (\theta, \varphi, t) \equiv
\textrm{Tr} [\hat J^+(\Delta\mathbf k) \hat J^-(\Delta\mathbf k)
\rho(t)] $ is the collective factor where $\rho(t)$ is the atomic
density matrix. At the initial time of photon emission,
\begin{eqnarray}
 I_c (\theta, \varphi, 0) &=& \langle \hat{N}_s \rangle + \sum_{j\ne
    j'}  \mathrm e^{- i\Delta\mathbf k\cdot (\mathbf
    r_j -\mathbf r_{j'})} \langle \hat \sigma_{j'}^+ \hat \sigma_j^-  \rangle   \label{pattern}   \\
& = & \langle \hat{N}_s \rangle - \frac{P}{N-1} +P  \frac{ |\langle \sum_j e^{- i\Delta\mathbf k\cdot \mathbf r_j}  \rangle|^2}{N^2-N},  \notag
\end{eqnarray}
where $ \hat{N}_s \equiv \sum_j (\hat \sigma_j^z +1)/2 $ is the
spin-excitation number operator. Here and hereafter $\langle \cdots
\rangle$ denotes the expectation value over $\rho(0)$, the initial
many-body state of interest. $P \equiv \langle \sum_{j \neq j'} \hat
\sigma_{j'}^+ \hat \sigma_j^- \rangle $ is the sum of spin
pair-correlations. The last equal sign in Eq.~(\ref{pattern}) holds
when $\rho (0)$ is invariant under permutation of atoms, which is the
typical situation for atom gases. $|\langle \sum_j e^{- i\Delta\mathbf
  k\cdot \mathbf r_j} \rangle|^2 $ is a sharp feature which equals
$N^2$ along the forward direction ($\theta=0$), and drops to zero for
$\theta \geq \theta_b \equiv \min \{ \sqrt{\frac{\pi}{k_0 H} },
\frac{2 \pi }{k_0 A} \}$ where $A $ and $H$ are respectively the
transverse and longitudinal size of the ensemble~(Fig.~\ref{fig1}). Thus,
positive (negative) pair-correlation sum manifests as a sharp
diffraction peak (dip), and its magnitude can be read out from the
ratio of the peak (dip) to the background: 
\begin{eqnarray}
  \frac{I(\theta=0) - I
    (\theta_b)}{I (\theta_b)} =\frac{P}{\langle \hat{N}_s \rangle -
    P/N}.
  \label{ratio}
\end{eqnarray}

For general states in optical lattices without the permutation
symmetry, $P$ can be measured after sudden release of atoms into a
spin-independent trap~\cite{Bloch_RMP}. The density matrix averaged
over many ensemble copies will become permutation-symmetric after
atoms lose memory of their initial positions, while $P$ is preserved
by the atomic motions. Moreover, we find that pair-correlation sum of
a dilute hot atomic vapor can be measured in the same way if Stokes
photon emission is controlled to be much slower than atomic motions
(see last part of Appendix \ref{perturbation}).

\section{Entanglement Detection}

The pair-correlation sum measured from the peak (dip) to background
ratio (Eq.~(\ref{ratio})) can detect entanglement via spin squeezing
inequalities~\cite{Toth_squeezing,Duan2011,Sorensen_spinsqueezing,Korbicz_spinsqueezing}.
The longitudinal component of total spin is equivalent to the
spin-excitation number: $ \hat{N}_s \equiv \hat J_z + N/2$, and the
second moment of transverse components is equivalent to the
pair-correlation sum: $\langle \hat J_x^2 \rangle + \langle \hat J_y^2
\rangle = P + N/2$. Many spin squeezing inequalities derived for first
and second moments of total spin can thus be formulated as
pair-correlation sum rules. For example, the optimal spin squeezing
inequalities discovered in Ref.~\cite{Toth_squeezing} become:
\begin{subequations}
\begin{eqnarray}
P  &\leq&  (N-1)\Delta N_s ^2   , \label{witness1}\\
P  &\geq&  -\Delta N_s ^2  , \label{witness2}\\
(N-1) P &\geq& \langle \hat{N}_s ^2 \rangle - N  \langle \hat{N}_s \rangle.
\label{witness3}
\end{eqnarray}
\label{entangle_witness}
\end{subequations}
Where $\Delta N_s \equiv (\langle \hat{N}_s^2 \rangle - \langle
\hat{N}_s \rangle^2)^{1/2}$. Violation of any one of the inequalities
(\ref{witness1}-\ref{witness3}) implies entanglement. With the
spin-excitation number $\hat{N}_s$ conserved in most physical
processes of interest, its expectation value are usually known
\textit{a priori}.

$\Delta N_s$ can also be measured from the peak (dip) to background
ratio in the diffraction image taken after a global rotation of the
ensemble. With a $\pi/2$ about an in-plane axis transforming $\hat
J^x\to\hat J^z$ or $\hat J^y\to\hat J^z$, $\langle \hat J_y^2 \rangle
+ \langle \hat J_z^2 \rangle - N/2$ or $\langle \hat J_x^2 \rangle +
\langle \hat J_z^2 \rangle - N/2$ can be obtained from the peak (dip)
to background ratio in the diffraction image, from which we can solve for $\Delta N_s$.

Entanglement detection based on the above pair-correlation sum rules
is described by the phase diagrams shown in Fig.~\ref{fig2} (a-c).
\textit{Qualitative} criteria become possible for entanglement witness
in two limits.  With vanishing $\Delta N_s$ seeing either a
diffraction peak or dip verifies entanglement, while with maximum
$\Delta N_s$ seeing a dip verifies entanglement (Fig.~\ref{fig2}
(a-b)). On the other hand, a peak (dip) strength exceeding some
threshold value always implies entanglement. Taking
half-spin-excitation states for example, observing a dip to background
ratio $|r| \geq \frac{1}{2}$ or a peak to background ratio $r \geq
\frac{N(N-1)}{N+1}$ verifies entanglement for any possible $\Delta
N_s$. $P$ and $\Delta N_s$ can also quantify the entanglement depth in
the vicinity of Dicke states~\cite{Duan2011}.

Furthermore, the diffraction image can be used to measure delocalized
entanglement as defined in Ref.~\cite{{Delocalized_entanglement}} for
atoms in optical lattices. A measure of the bipartite delocalized
entanglement at specified distance $\mathbf x$ is given by the
entanglement of formation for delocalized bipartite reduced density
operator
\begin{eqnarray}
  \rho_{AB}(\mathbf x)\equiv\frac{1}{C_{\mathbf
      x}}\sum_\mathbf j\rho_{\mathbf j,\mathbf j+\mathbf x}.
\end{eqnarray}
Here $C_{\mathbf x}$ is the normalization coefficient which
corresponds to the number of pairs $\{\mathbf j,\mathbf j+\mathbf
x\}$. $\rho_{\mathbf j,\mathbf j+\mathbf x}$ denotes the two-qubit
reduced density matrix deduced from the initial ensemble state
$\rho(0)$, where only the sites $\mathbf j$ and $\mathbf j+\mathbf x$
of the lattice are kept while all others are traced out.

As shown in Ref.~\cite{{Delocalized_entanglement}}, the lower bound of
entanglement of formation for $\rho_{AB}(\mathbf x)$ can be evaluated
from the fidelity $f_{\phi}(\mathbf
x)\equiv\langle\phi|\rho_{AB}(\mathbf x)|\phi\rangle$, with $\phi$
being one of the four Bell states $\Phi_\pm$ and $\Psi_\pm$. The
fidelity is found to be
\begin{eqnarray}
  f_{\Phi_\pm}(\mathbf x)=\frac{1-\textrm{Tr}[ \sum_{\mathbf j}
    \hat\sigma_{\mathbf j}^z \hat\sigma_{\mathbf j+ \mathbf x}^z
    \rho(0)]}{4}\pm\frac{P_{\mathbf x}+P_{-\mathbf x}}{2},
\end{eqnarray}
where the correlation $P_{\mathbf x} \equiv\textrm{Tr}[ \sum_{\mathbf
  j} \hat\sigma_{\mathbf j}^+ \hat\sigma_{\mathbf j+ \mathbf x}^-
\rho(0)]$. $f_{\Psi_\pm}(\mathbf x)$ can be obtained from
$f_{\Phi_\pm}(\mathbf x)$ by applying a global unitary transformation.

Eq.~(\ref{pattern}) can be rewritten as
$I_c(\theta,\varphi,0)=\langle\hat{N}_s\rangle+\sum_{\mathbf x}\mathrm
e^{- i\Delta\mathbf k\cdot\mathbf x} C_{\mathbf x} P_{\mathbf{x}}$.
The correlation $P_{\mathbf{x}}$ for arbitrary $\mathbf x$ can
therefore be obtained through a Fourier transform of the diffraction
image. Note that $P_{\mathbf x}+P_{-\mathbf
  x}=\textrm{Tr}[\sum_{\mathbf j} (\hat\sigma_{\mathbf
  j}^x\hat\sigma_{\mathbf j+ \mathbf x}^x+\hat\sigma_{\mathbf
  j}^y\hat\sigma_{\mathbf j+ \mathbf x}^y)\rho(0)]$. Thus,
$\textrm{Tr}[\sum_{\mathbf j} \hat\sigma_{\mathbf j}^z
\hat\sigma_{\mathbf j+ \mathbf x}^z \rho(0)]$ can also be obtained by
applying a global rotation to all spins to transform
$\hat\sigma^x\to\hat\sigma^z$ or $\hat\sigma^y\to\hat\sigma^z$.

\section{Perturbative Solution of the Atomic Evolution}

Hereafter, we focus on dilute ensembles where interatomic distance is
comparable to or larger than optical wavelength.  Remarkably, under
this condition, one can collect all Stokes photons, not only those
initial ones, for measuring the pair-correlation sum and detect
entanglement in $\rho(0)$. 

The diffraction pattern at an arbitrary time is determined by the
instantaneous atomic density matrix $\rho(t)$ which differs from
$\rho(0)$. As well
established in the literature of
superradiance~\cite{Carmichael_superradiance}, the evolution of $\rho(t)$ is described by the
Lindblad master equation in the Born-Markov approximation, 
\begin{eqnarray}
  \dot{\rho}(t)&=&\mathcal{L}_0\rho(t)+\mathcal{L}_1\rho(t), \label{original_ME} \\
  \mathcal{L}_0\rho &\equiv& \frac{\Gamma}{2}\sum_j(2\hat\sigma_j^-\rho\hat\sigma_j^+-\hat\sigma_j^+\hat\sigma_j^-\rho-\rho\hat\sigma_j^+\hat\sigma_j^-) \notag \\
  \mathcal{L}_1\rho &\equiv & \sum_{j\ne
    j'}\frac{\Gamma_{jj'}}{2}\left(2\hat\sigma_j^-\rho\hat\sigma_{j'}^+-\hat\sigma_{j'}^+\hat\sigma_j^-\rho-\rho\hat\sigma_{j'}^+\hat\sigma_j^-\right)\notag \\
  & & +i\sum_{j\ne j'}\frac{G_{jj'}}{2}[\hat\sigma_{j'}^+\hat\sigma_j^-,\rho], \notag
\end{eqnarray}
where $\Gamma_{j j'}=\Gamma\frac{\sin(k_0 |\mathbf r_{j}-\mathbf
  r_{j'}|)}{k_0 |\mathbf r_{j}-\mathbf r_{j'}|} $ and
$G_{jj'}=\Gamma\frac{\cos(k_0|\mathbf r_{j}-\mathbf r_{j'}|)}{k_0
  |\mathbf r_{j}-\mathbf r_{j'}|}$ describe respectively the multiple
light scattering and dipole-dipole
interaction~\cite{Wstate,Carmichael_superradiance}. In the study of
super-radiance phenomena in very dense atomic ensembles, these effects
must be accounted
non-perturbatively~\cite{Eberly_superradiance}. $\Gamma_{j j'}$ and
$G_{jj'}$ drop fast with distance. In dilute atomic ensembles where
the atom-atom distance is comparable or larger than the photon
wavelength, the atomic evolution can be solved perturbatively.  Using
the Laplace transform $w(z)=\int_0^{\infty}dt\mathrm e^{-zt}\rho(t)$,
we have
\begin{align}
  w(z)=\frac{1}{z-\mathcal{L}_0-\mathcal{L}_1}\rho(0). \notag
\end{align}
For $k_0 |\mathbf r_{j}-\mathbf r_{j'}| \geq 2 \pi$, $\mathcal{L}_1$
is small compared to $\mathcal{L}_0$, and we make a perturbative
expansion
$\frac{1}{z-\mathcal{L}_0-\mathcal{L}_1}=\frac{1}{z-\mathcal{L}_0}+\frac{1}{z-\mathcal{L}_0}\mathcal{L}_1\frac{1}{z-\mathcal{L}_0}+\cdots$. By
inverse Laplace transform we can get the solution of the atomic
density matrix $\rho^{(n)}$ keeping up to the $n$-th order effects of
$\mathcal{L}_1$.

For the zeroth order solution $\rho^{(0)}(t)=\mathrm
e^{\mathcal{L}_0t}\rho(0)$, we find $ \textrm{Tr} [\hat J^+(\Delta\mathbf
k)\hat J^-(\Delta\mathbf k) \rho^{(0)}(t)] = e^{- \Gamma t} \textrm{Tr}
[\hat J^+(\Delta\mathbf k)\hat J^-(\Delta\mathbf k) \rho(0)] $, i.e. the initial
diffraction pattern is preserved for all time. Comparisons with exact
solution of master equation for a chain of $12$ atoms show that the
perturbation expansion converges fast for $k_0 |\mathbf r_{j}-\mathbf
r_{j'}| \geq 2 \pi$, and the effects of the multiple light scattering
and dipole-dipole interaction are well accounted by keeping only the
first order effects of the $\mathcal{L}_1$ term:
\begin{align}
  &\rho^{(1)}(t)=\rho^{(0)}(t)+\int_0^td\tau\mathrm
  e^{\mathcal{L}_0\tau}\Big[\mathcal{L}_1\rho^{(0)}(t-\tau)\Big] \notag.
\end{align}
The $\mathcal{L}_1$ term leads to slow varying modulation of the
diffraction pattern, which barely changes the ratio of the sharp peak
(dip) to its neighboring background. Details on this modulation and
the convergence check for the perturbative solutions can be found in
Appendix~\ref{perturbation}.

\begin{figure*}[tbp]
\includegraphics[width=12cm]{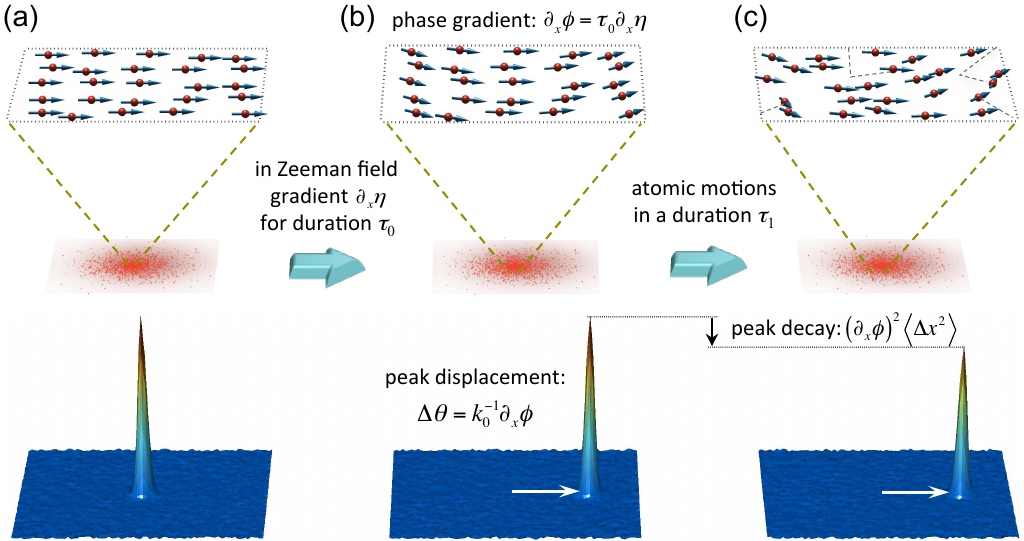}
\caption{\label{Fig3} Diffraction images (lower parts) from atomic
ensemble in different spin configurations (upper). (a)
Spin-coherent-state with in-plane polarization. (b) Evolution in a
Zeeman field gradient imprints a phase gradient of spins, resulting
in a displacement of the diffraction peak which can be a principle
of gradiometer. (c) Atomic motions diminish the spin polarization,
resulting in decay of the displaced peak. This can be a principle
for non-demolition measurement of atomic temperature and collisional
interactions. \label{fig3}}
\end{figure*}

Based on this perturbative solution, we analyze the diffraction
pattern of Stokes photons as a function of the collection time
$\tau_c$. We find the weak processes of $\mathcal{L}_1$ only result in
slow varying modulation of the diffraction pattern, barely changing
the ratio of the sharp peak or dip to its neighboring
background. Namely,
\begin{equation}
  r(\tau_c) \equiv \frac{n(\theta=0, \tau_c) - n(\theta_b, \tau_c)}{n(\theta_b, \tau_c)} \cong \frac{P}{\langle \hat{N}_s \rangle - P/N} \label{longtime}
\end{equation}
where $n(\theta, \varphi, \tau_c) \equiv \delta\Omega \int_0^{\tau_c}
dt I (\theta, \varphi, t)$ is the number of photons emitted into an
infinitesimal solid angle in the collection interval $0 \leq t \leq
\tau_c$. For a dilute ensemble Eq.~(\ref{longtime}) holds for
arbitrarily large $\tau_c$ (cf. Fig.~\ref{fig2} (d)). Thus the pair-correlation
sum can be faithfully read out from the diffraction pattern of all
photons, not limited to those initial ones.

We estimate the range of applicability of our treatment. The dilute
condition is satisfied by typical cold atom gases of a density
$10^{10} - 10^{12}$~cm$^{-3}$ or by atoms in optical lattice. The
duration of Stokes photon emission is of a timescale $ \Gamma^{-1} =
(\frac{\Omega_L}{2\Delta})^{-2} \Gamma_0^{-1}$. The excited state
decay rate $\Gamma_0 \gtrsim 30~\mu$s$^{-1}$ for typical alkali
atoms.~\cite{quantum_memory0,quantum_memory1,quantum_memory2,quantum_memory3,quantum_memory4,Pan_quantummemory,dephaseinOL1,dephaseinOL2}
Taking $(\frac{\Omega_L}{2\Delta})^{-2} \sim 40$, all Stokes photons
are emitted in a timescale $\Gamma^{-1} \lesssim ~\mu$s. For cold atom
gases with a temperature of $1 - 100~\mu$K, the average velocity is $
0.01 - 0.1$ m/s. Atoms can only travel $10 - 100$ nm in the duration
of $\Gamma^{-1}$ which is indeed negligible as compared to the light
wavelength.

\section{Field Gradiometer and non-demolition probe of atomic motions and temperature}

Under free evolution, the pair-correlation changes as $
\textrm{Tr}[\hat \sigma_{i'}^+\hat \sigma_i^- \rho(\tau)] =
e^{i(\eta_i \tau - \eta_{i'} \tau) - 2 \gamma \tau} \textrm{Tr}[\hat
\sigma_{i'}^+\hat \sigma_i^- \rho(0)] $, where $\eta_i$ is the Zeeman
frequency and $ \gamma$ the homogeneous dephasing rate of an
individual spin. The pair-correlation sum thus decays only at the
single spin dephasing rate. Therefore entanglement in $\rho(0)$ can be
reliably detected from the dephased state $\rho(\tau)$ as long as
$\tau \ll \gamma^{-1}$, even when the fidelity is exponentially small
with $N$~\cite{Duan2011}.

Spatial inhomogeneity of external fields leads to position dependent
Zeeman frequency $\eta(\mathbf{r})$ and hence inhomogeneous
precessions of spins. If the size of the ensemble is small compared to
the variation length scale of the field, the dominating term is the
gradient: $\eta(\mathbf{r}) \cong \mathbf{r} \cdot \mathbf{\nabla}
\eta $. For an ensemble initially in a permutation-symmetric state,
after an interval $\tau_0$ with frozen motion in the Zeeman field
gradient, the diffraction pattern becomes $I_c = \langle \hat{N}_s
\rangle -\frac{P}{N-1} + \frac{P}{N^2-N} |\langle \sum_j e^{-
  i(\Delta\mathbf k - \tau_0 \mathbf{\nabla} \eta)\cdot \mathbf r_j}
\rangle|^2$. We focus on situations where $\partial_z \eta $ is either
zero or not picked up by atomic ensembles of a quasi-2D geometry in
$x-y$ plane. The in-plane gradient simply results in a displacement of
the sharp diffraction peak or dip, preserving its strength and shape
(Fig.~\ref{fig3}). This has several significant consequences. First,
by evolution in an external field of known gradient, entanglement can
be detected by measuring the peak or dip along a chosen direction with
finite $\theta$, such that detectors do not pick up laser
photons. Second, the displacement measures the vector value of the
gradient.  It can thus be used as a principle of vector gradiometer of
magnetic field, static electric field via dc Stark effect, and light
field via ac Stark effect~\cite{dephaseinOL1}.

An ideal probe state is the spin-coherent-state of $N$ unentangled
atoms with in-plane polarization, which can be realized by optical
pumping followed by a spin rotation to the in-plane direction. The
gradient is then probed simultaneously by the $\sim N^2$ classical
pair-correlations, and its vector value is encoded as the displacement
of a diffraction peak with strength $\sim N^2$. 

\begin{figure*}[tbp]
\includegraphics[width=11cm]{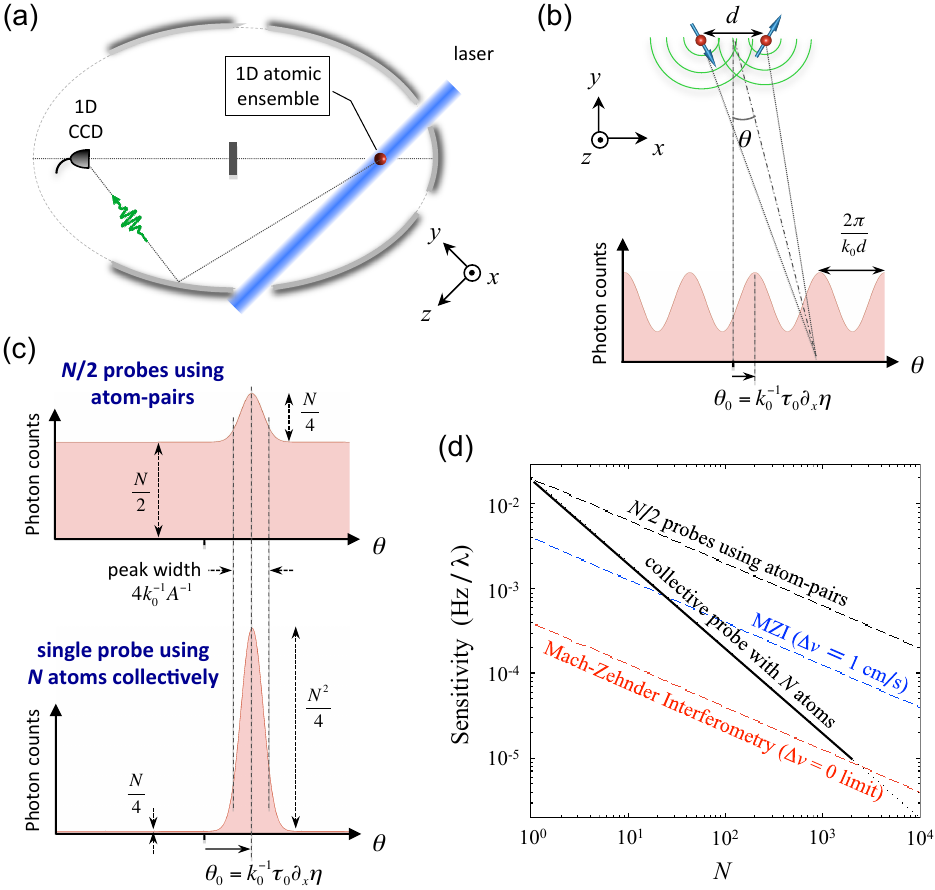}
\caption{\label{Fig4} Zeeman field gradiometer using a 1D atomic
  ensemble. (a) and (b) Schematic of the setup. The ensemble and a 1D
  CCD array are placed respectively on the two common focus lines of a
  group of elliptical cylinder mirrors. The mirrors ensure the majority of photons is collected by the detector. The smallest ensemble can be an atom-pair, giving a direct
  analog of the double-slit interferometry. (c) Signal of
  $\frac{N}{2}$ probes using atom-pairs (upper), and single probe
  using $N$ atoms collectively (lower), with each atom placed randomly
  on the focus line according to a Gaussian distribution with FWHM
  $A$. The peak strength of the lower is enhanced by a factor of
  $N$. (d) Gradiometer sensitivity at a spatial resolution $A=1$~mm
  with a resource of $N$ unentangled atoms. The diffraction based
  gradiometer using $N$ atoms collectively ($\frac{N}{2}$ atom-pairs
  independently) has a sensitivity of the $\frac{1}{N}$
  ($\frac{1}{\sqrt{N}}$) scaling, shown by the solid (dashed) black
  line. Sensitivity of flying atom Mach-Zehnder interferometry (MZI)
  gradiometer is shown for reference.~\cite{MZI_gradiometer1} The
  probe time $\tau_0=\frac{A}{\Delta v}= 0.1$~s for the blue line,
  limited by a finite velocity uncertainty $\Delta v = 1~$cm/s, while
  $\tau_0=1$~s for all other lines, limited only by the single spin
  dephasing time. \label{fig4}}
\end{figure*}

Now we analyze the sensitivity of our diffraction based Zeeman field
gradiometer. Consider the atoms of a 1D geometry illustrated in
Fig.~\ref{fig4} (a) with a Gaussian spatial distribution of
full-width-half-maximum (FWHM) $A$. Atoms are initialized in the
spin-coherent-state and evolved in the Zeeman field gradient for an
interval of $\tau_0$. The diffraction pattern is then:
$\frac{N^2}{4}f(\theta)+\frac{N}{4}$, where $f(\theta)\equiv\mathrm
e^{-\frac{(k_0A)^2}{4}(\theta-\theta_0)^2}$ is a sharp peak centered
in a tilted direction: $\theta_0=k_0^{-1} \tau_0 \partial_x \eta$. Our
goal is to extract this direction from the photon statistics. The
field gradient can then be inferred based on the above relation. The
spatial resolution of the gradiometer is just given by the size of the
atomic ensemble $A$. The precision of this measurement is determined
by the width of the peak ($\frac{1}{k_0A}$), the shot noise of the
photon counts and the angular resolution ($\delta\theta$) of the CCD
detector array. While the CCD angular resolution can always be
improved by increasing the distance from the atomic ensemble, the
former two factors will determine the quantum limit for the
sensitivity of this gradiometer. We will examine the increase of the
sensitivity with number of atoms $N$ used in the probe. Our discussion
is limited to the dilute regime (i.e. $k_0 A / N \geq 2 \pi$).

In a single probe using $N$ atoms, the photon counts at each CCD pixel
can be written as $n_i + \Delta n_i$, where $n_i $ and $ \Delta n_i$
are respectively the expectation value and fluctuation of the photon
counts. We have
\begin{eqnarray}
  n_i&=&\int_{\theta_i-\frac{\delta\theta}{2}}^{\theta_i+\frac{\delta\theta}{2}}d\theta
  \bigg(\frac{N^2}{4}f(\theta)+\frac{N}{4}\bigg)\nonumber\\
  &=&\delta\theta \frac{N^2}{4}\bar f(\theta_i)+\delta\theta\frac{N}{4} , \nonumber\\
 \end{eqnarray}
where $\bar f(\theta_i)\equiv\frac{1}{\delta\theta}\int_{\theta_i-\frac{\delta\theta}{2}}^{\theta_i+\frac{\delta\theta}{2}}d\theta f(\theta)$. Here we assume that the $i$-th pixel of the detector collects all photons emitted within the angle range
$[\theta_i-\frac{\delta\theta}{2},\theta_i+\frac{\delta\theta}{2}]$, where $\theta_i\equiv i\delta\theta$. As shown in
Appendix~\ref{fluctuation}, the photon statistics is found to be
Poissonian when the probe state is spin-coherent-state, and we have
\begin{equation}
 \langle\Delta n^2_i\rangle \sim n_i,
\end{equation}

From the photon statistics $\{ n_i + \Delta n_i\}$, we can extract a
peak central position $\theta_c$ defined as
\begin{eqnarray}
  \theta_c=\frac{\sum_i\theta_i(n_i+\Delta n_i)}{\sum_i(n_i+\Delta
    n_i)}. \label{peakcenter}
\end{eqnarray}
$\theta_c$ unavoidably has some deviation from $\theta_0$, the peak
position precision is then defined as $\sqrt{(\theta_c
  -\theta_0)^2}$. Our analysis shows that (see Appendix
\ref{sen_diffract}), when $\theta_c$ from a single probe is used to
extract the Zeeman field gradient $\partial_x \eta$, the overall
precision is
\begin{equation} 
  \Delta (\partial_x \eta) \sim \frac{k_0}{\tau_0} \sqrt{\frac{4\pi^{-1/2}}{N^2k_0A}
    +(k_0A)^4\delta\theta^6}
\end{equation}
For small $\delta\theta$, the sensitivity is $\Delta (\partial_x \eta)
\sim \frac{k_0}{\tau_0} \frac{1}{N\sqrt{k_0A}}$ which
scales inversely with $N$.

The smallest ensemble for the gradiometer can just be an atom pair
prepared on spin-coherent-state which emit one photon on average in
each probe. We can make $N/2$ independent probes using such atom pairs
and extract the field gradient from the integrated signal. For one
atom at position $\mathbf x_1$ and the other atom at $\mathbf x_2$,
the photon has an emission distribution of
$\sim1+\frac{1}{2}\cos\mathbf k\cdot(\mathbf x_1-\mathbf x_2)$ (see
Fig.~\ref{fig4} (b)). This is a direct analog of the double-slit
interferometry. We assume each atom has fixed position during a single
probe but it can randomly appear in a 1D line according to a Gaussian
distribution $\propto\mathrm e^{-2x^2/A^2}$ with
full-width-half-maximum $A$ for multiple probes. Summing over $N/2$
probes, the total photon distribution pattern is
$\frac{N}{2}+\frac{N}{4}\mathrm e^{-(k_xA)^2/4}$. The peak strength is
thus $\sim N$. Following the previous derivations, we find sensitivity
of measuring the Zeeman field gradient is
$\frac{k_0}{\tau_0}\frac{1}{\sqrt{Nk_0A}}$ with the $1/\sqrt{N}$
scaling. 

Thus, a single collective probe using $N$ atoms has sensitivity $\sim
\frac{1}{N} \frac{k_0}{ \tau_0 \sqrt{k_0 A}}$, which goes beyond the
SQL of $\sim \frac{1}{\sqrt{N}}\frac{k_0}{ \tau_0 \sqrt{k_0 A}}$ for
$N/2$ independent probes using atom-pairs, as shown in Fig.~\ref{fig4}
(d).  The enhancement comes from the $N^2$ scaling of the peak
strength, which is the result of using large group of atoms
collectively. In Fig.~\ref{fig4} (d), we also compare with the
gradiometer based on Mach-Zehnder interferometer (MZI) of flying atoms
in atomic fountain which has the SQL sensitivity of $\sim
\frac{1}{\sqrt{N}} \frac{1}{ \tau_0 A}$, while the inevitable velocity
uncertainty further sets a tighter upper bound for $\tau_0$ dependent
on the spatial resolution $A$ (see Appendix \ref{sen_MZI}).

The collectively-enhanced sensitivity with $1/N$ scaling is valid in
the dilute regime $N \leq k_0 A$. Beyond this regime, the multiple
light scattering can not be treated perturbatively and its effect will
eventually renormalize the peak strength to the $N$ scaling. By
trapping atoms in optical lattices, the probe time $\tau_0$ can be as
long as the single spin homogeneous dephasing time, in the order of
second or longer~\cite{dephaseinOL1}. This diffraction based
gradiometer using stationary atoms is immune to collective noises and
uncertainty in atomic positions, and can have a fine spatial
resolution (given by the size of the ensemble). Remarkably, for the
scheme to work, the probe state does not need to have high degree of
spin polarization as an imperfect polarization $p$ just scales down
the sensitivity by $1/p$.

The diffraction image can also be used for non-demolition probe of
atomic motions and temperature in trapped cold atom gases, by
introducing a waiting time $\tau_1$ between the imprinting of phase
gradient $\nabla \phi$ on the spin-coherent-state and the measurement
of the Stokes photon diffraction (Fig.~\ref{fig3}). Atomic motions in
the interval $\tau_1$ will diminish the spin polarization, resulting
in decay of the displaced diffraction
peak~\cite{Pan_quantummemory}. For $| \nabla \phi |^2 \langle \Delta
\mathbf r^2 \rangle \ll1$, the peak strength is $\frac{N^2}{4}e^{ -
  |\nabla \phi |^2 \langle \Delta \mathbf r^2 \rangle /3 } $, $
\langle \Delta \mathbf r^2 \rangle$ being the mean square displacement
of atoms. For short $\tau_1$ when $\Delta \mathbf r$ is small compared
to the interatomic distance $d$, $ \langle \Delta \mathbf r^2 \rangle
=2 \frac{k_B T}{m} \tau_1^2$.  Thus, by preparing a large phase
gradient $|\nabla \phi | \sim 1/d$, the short time motion can be
probed and the atomic temperature can be read out from the decay of
the peak.  Smaller $|\nabla \phi |$ allows the probe of long time
motion which will eventually crossover to the diffusive regime by atom
collisions. $\tau_1$ is upper limited by the spin dephasing time,
which is long enough for observing the entire crossover behavior from
ballistic to diffusive motions, providing information about the
collisional interactions in trapped gases. The collectively enhanced
peak strength of $\sim N^2$ provides sufficient signal-to-noise ratio
for determining $ \langle \Delta \mathbf r^2 \rangle$ at a given
$\tau_1$ by a single shot measurement.

\section{Summary}

In conclusion, we have shown that the far field diffraction image of spontaneously
emitted Raman photons can be used for detection of spin entanglement
in cold atomic ensembles as well as for quantum metrology applications. 
For many-body states with small or maximum uncertainty in spin-excitation number, entanglement is witnessed by the presence of either a sharp diffraction peak or dip. For general states, the relative strength of the peak or dip over its background detects entanglement through the pair-correlation sum rules derived from spin squeezing inequalities. 
Spin precessions in Zeeman field gradient lead to displacement of the diffraction peak or dip while atomic motions lead to decay of its strength. These can serve as principles for vector gradiometer of fields and for non-demolition measurement of atomic temperature and collisional dynamics. The gradiometer sensitivity
can reach $1/N$ by using a spin-coherent-state of $N$ unentangled
atoms as the probe, which suggests a new possibility for going beyond
the SQL without entanglement.  Motional dynamics leads to temporal
decay of the diffraction peak which can be used for non-demolition
probe of temperature and collisional interactions in trapped atomic
gases.

\section{Acknowledgments}

The authors thank Lian-ao Wu for helpful discussions. WY thanks CQI at IIIS of Tsinghua for hospitality during his visit
through the support by NBRPC under grant 2011CBA00300
(2011CBA00301). The work was supported by the Research Grant Council
of Hong Kong under grant HKU706711P and HKU8/CRF/11G.

\appendix

\section{Perturbative Solution to the Master Equation}
\label{perturbation}

We assume the wave vector $\mathbf k_0$ of the driving laser is
perpendicular to the ensemble (i.e.  $\mathbf k_0\cdot\mathbf r_i=0$).
For the simplicity of expression, below we replace $\Delta\mathbf
k\equiv\mathbf k-\mathbf k_0$ by $\mathbf k$ when it appears in $\hat
J^-(\Delta\mathbf k)$.

The collected photon number in a time $\tau_c$ along direction
$\mathbf k$ within the infinitesimal solid angle $\delta\Omega$ writes
$n_p(\mathbf
k,\tau_c)=\Gamma\frac{\delta\Omega}{4\pi}\int_0^{\tau_c}dt\textrm{Tr}[\hat
J^+(\mathbf k)\hat J^-(\mathbf k)\rho(t)]$. Here we have ignored the
slowly varying single atom dipole emission pattern. $\rho(t)$ can be
solved from the master equation (Eq. (\ref{original_ME})) by viewing
the $\mathcal{L}_1$ term as perturbation.

We use the notation $\rho^{(n)}(t)$ and
$\langle\cdots\rangle^{(n)}_t\equiv\textrm{Tr}[\cdots\rho^{(n)}(t)]$
to describe the result keeping up to $n$-th order effect of
$\mathcal{L}_1$. For the $0$-th order result $\rho^{(0)}(t)=\mathrm
e^{\mathcal{L}_0t}\rho(0)$, there is
$\langle\hat\sigma_m^+\hat\sigma_n^-\rangle^{(0)}_t=\mathrm e^{-\Gamma
  t}\langle\hat\sigma_m^+\hat\sigma_n^-\rangle_0$. Then
\begin{align}
n_p^{(0)}(\mathbf k,\tau_c)=\frac{\delta\Omega}{4\pi}(1-\mathrm e^{-\Gamma\tau_c})\langle\hat
J^+(\mathbf k)\hat J^-(\mathbf k)\rangle_0.  \label{0th}
\end{align}

Below we solve for $\rho^{(1)}(t)$ which captures the leading order
effect of $\mathcal{L}_1$, and show that it is sufficient to account
for the effect of $\mathcal L_1$ under the dilute limit. We consider
an $N$-atom 1D lattice. In the 1-st order approximation, the equation
of motion for the pair-correlation is
\begin{align}
  \frac{d\langle\hat\sigma_m^+\hat\sigma_n^-\rangle^{(1)}_t}{d(\Gamma t)}
  =&-\langle\hat\sigma_m^+\hat\sigma_n^-\rangle^{(1)}_t\nonumber\\
  &+\sum_{j\ne
    n}(\frac{\Gamma_{jn}}{2\Gamma}-i\frac{G_{jn}}{2\Gamma})\langle\hat
  \sigma_m^+\hat\sigma_j^-\hat\sigma_n^z\rangle^{(0)}_t\nonumber\\
  &+\sum_{j\ne
    m}(\frac{\Gamma_{jm}}{2\Gamma}+i\frac{G_{jm}}{2\Gamma})\langle\hat
  \sigma_m^z\hat\sigma_j^+\hat\sigma_n^-\rangle^{(0)}_t.
  \label{mn_diffeq}
\end{align}
We denote $\Gamma_j\equiv\Gamma_{n\pm
  j,n}=\Gamma\frac{\sin|jkd|}{|jkd|}$, $G_j\equiv G_{n\pm
  j,n}=\Gamma\frac{\cos|jkd|}{|jkd|}$, and
$a^{(0)}_t\equiv\langle\hat\sigma_m^+\hat\sigma_m^-\rangle^{(0)}_t$ ,
$p^{(0)}_t\equiv\langle\hat\sigma_m^+\hat\sigma_n^-\rangle^{(0)}_{t,m\ne
  n}$. Then
\begin{align}
  \frac{d\langle\hat\sigma_n^+\hat\sigma_n^-\rangle^{(1)}_t}{d(\Gamma t)}=&-\langle\hat
  \sigma_n^+\hat\sigma_n^-\rangle^{(1)}_t-p^{(0)}_t\Big(\sum_{j=1}^{N-n}+\sum_{j=1}^{n-1}\Big)
  \frac{\Gamma_j}{\Gamma},\nonumber\\
  \frac{d\langle\hat\sigma_m^+\hat\sigma_n^-\rangle^{(1)}_{t,m\ne
      n}}{d(\Gamma t)}=&-\langle\hat\sigma_m^+\hat\sigma_n^-\rangle^{(1)}_{t,m\ne
    n}+2B^{(0)}_t\frac{\Gamma_{m-n}}{\Gamma}\nonumber\\
  &+\alpha^{(0)}_t\Big(\sum_{j=1}^{N-n}+\sum_{j=1}^{n-1}\Big)
  \big(\frac{\Gamma_j}{\Gamma}-i\frac{G_j}{\Gamma}\big)\nonumber\\
  &+\alpha^{(0)}_t\Big(\sum_{j=1}^{N-m}+\sum_{j=1}^{m-1}\Big)
  \big(\frac{\Gamma_j}{\Gamma}+i\frac{G_j}{\Gamma}\big).
  \label{mn}
\end{align}
Where
$\alpha^{(0)}_t\equiv\frac{1}{2}\langle\hat\sigma_j^+\hat\sigma_m^-\hat
\sigma_n^z\rangle^{(0)}_{t,j\ne m\ne n}$ is the three-body
correlation,
$B^{(0)}_t\equiv\frac{1}{2}\langle\hat\sigma_m^+\hat\sigma_m^-\hat\sigma_n^z\rangle^{(0)}_{t,m\ne
  n}-\alpha^{(0)}_t$.

Using $\langle\hat J^+(\mathbf k)\hat J^-(\mathbf
k)\rangle^{(1)}_t=\sum_n\langle\hat\sigma_n^+\hat
\sigma_n^-\rangle^{(1)}_t+\sum_{m\ne n}\mathrm e^{i\mathbf
  k\cdot(\mathbf r_m-\mathbf r_n)}\langle\hat\sigma_m^+\hat
\sigma_n^-\rangle^{(1)}_t$, and switching the summation index as
\begin{align}
  \sum_{n=1}^N\Big(\sum_{j=1}^{N-n}+\sum_{j=1}^{n-1}\Big)
  =\sum_{j=1}^{N-1}\Big(\sum_{n=1}^{N-j}+\sum_{n=j+1}^N\Big),
\end{align}
we write
\begin{align}
  &\frac{d\langle\hat J^+(\mathbf k)\hat J^-(\mathbf
    k)\rangle^{(1)}_t}{d(\Gamma t)}\nonumber\\
  =&-\langle\hat J^+(\mathbf k)\hat J^-(\mathbf k)\rangle^{(1)}_t\nonumber\\
  &+4(N-j)\sum_{j=1}^{N-1}\frac{\Gamma_j}{\Gamma}
  \big(B^{(0)}_t\cos(j\mathbf k\cdot\mathbf d)-\frac{1}{2}p^{(0)}_t-\alpha^{(0)}_t\big)\nonumber\\
  &+\alpha^{(0)}_t\Big[\sum_{j=1}^{N-1}\big(\frac{\Gamma_j}{\Gamma}-i\frac{G_j}{\Gamma}\big)
    f_j(\mathbf k)+\textrm{C.c.}\Big].
\label{1storder_diffeq}
\end{align}
Where $\mathbf d$ is the vector connecting neighboring atoms, and 
\begin{align}
  f_j(\mathbf k)\equiv&\bigg(\sum_{m=1}^N\mathrm e^{i\mathbf
    k\cdot\mathbf
    r_m}\bigg)\bigg(\sum_{n=1}^{N-j}+\sum_{n=j+1}^N\bigg)\mathrm
  e^{-i\mathbf k\cdot\mathbf r_n}\nonumber\\
  =&2\cos^2\frac{j\mathbf k\cdot\mathbf
    d}{2}\frac{\sin^2\frac{N\mathbf k\cdot\mathbf
      d}{2}}{\sin^2\frac{\mathbf k\cdot\mathbf
      d}{2}}-\frac{\sin(N\mathbf k\cdot\mathbf d)\sin(j\mathbf
    k\cdot\mathbf d)}{2\sin^2\frac{\mathbf k\cdot\mathbf d}{2}}.
  \label{term_with_given_j}
\end{align}

As $\Gamma_j, G_j\sim\frac{1}{jkd}$, only those $j$ terms with $j\ll N$
  make significant contributions. Thus $\frac{\sin(N\mathbf
    k\cdot\mathbf d)\sin(j\mathbf k\cdot\mathbf
    d)}{2\sin^2\frac{\mathbf k\cdot\mathbf d}{2}}$ can be ignored
  compared to $\cos^2\frac{j\mathbf k\cdot\mathbf
    d}{2}\frac{\sin^2\frac{N\mathbf k\cdot\mathbf
      d}{2}}{\sin^2\frac{\mathbf k\cdot\mathbf d}{2}}$. Then
\begin{align}
  &\frac{d\langle\hat J^+(\mathbf k)\hat J^-(\mathbf
    k)\rangle^{(1)}_t}{d(\Gamma t)}\nonumber\\
  \approx&-\langle\hat J^+(\mathbf k)\hat
  J^-(\mathbf k)\rangle^{(1)}_t\nonumber\\
  &+\frac{2\alpha^{(0)}_t}{p^{(0)}_t}\sum_{j>0}
  \frac{\Gamma_j}{\Gamma}\big(\cos(j\mathbf k\cdot\mathbf
  d)+1\big)\langle\hat J^+(\mathbf k)\hat J^-(\mathbf
  k)\rangle^{(0)}_t \nonumber\\
  &+N\sum_{j>0}\frac{\Gamma_j}{\Gamma}\Big(C_2\cos(j\mathbf
  k\cdot\mathbf d)+C_3\Big), \label{diff}
\end{align}
with
$C_2\equiv4B^{(0)}_t-2\alpha^{(0)}_t(\frac{a^{(0)}_t}{p^{(0)}_t}-1)$
and
$C_3\equiv-2p^{(0)}_t-2\alpha^{(0)}_t(\frac{a^{(0)}_t}{p^{(0)}_t}+1)$.
In the above equation we have used the relation $\langle\hat
J^+(\mathbf k)\hat J^-(\mathbf
k)\rangle^{(0)}_t=N(a^{(0)}_t-p^{(0)}_t)+p^{(0)}_t\frac{\sin^2\frac{N\mathbf
    k\cdot\mathbf d}{2}}{\sin^2\frac{\mathbf k\cdot\mathbf d}{2}}$.

The time dependence of $a^{(0)}_t$, $p^{(0)}_t$, $\alpha^{(0)}_t$ and
$B^{(0)}_t$ are easily obtained from $\rho^{(0)}(t)$:
\begin{align}
  a^{(0)}_t=&~\mathrm e^{-\Gamma t}a^{(0)}_0,~~~~~~
  p^{(0)}_t=\mathrm e^{-\Gamma t}p^{(0)}_0, \nonumber \\
  \alpha^{(0)}_t=&~\mathrm e^{-2\Gamma
    t}\alpha^{(0)}_0-\frac{1}{2}\mathrm e^{-\Gamma t}(1-\mathrm
  e^{-\Gamma t})p^{(0)}_0,\nonumber\\
  B^{(0)}_t=&~\frac{1}{4}\mathrm e^{-2\Gamma t}\big\langle(\hat
  \sigma_m^z+1)(\hat
  \sigma_n^z+1)\big\rangle_0\nonumber\\
  &~-\frac{1}{2}\mathrm e^{-\Gamma t}a^{(0)}_0-\alpha^{(0)}_t.
  \label{time_evolution}
\end{align}

Solving the differential equation Eq.~(\ref{diff}), we obtain the photon
diffraction pattern under the $1$-st order approximation:
\begin{align}
  &n_p^{(1)}(\mathbf
  k,\tau_c)=\Gamma\frac{\delta\Omega}{4\pi}\int_0^{\tau_c}dt\langle\hat
  J^+(\mathbf k)\hat J^-(\mathbf k)\rangle_t \nonumber\\
  \approx&~\frac{\delta\Omega}{4\pi}N\sum_{j>0}\frac{\Gamma_j}{\Gamma}\bigg[f_2(\tau_c)\cos(j\mathbf
  k\cdot\mathbf d)+f_3(\tau_c)\bigg] \nonumber\\
  &~+\frac{\delta\Omega}{4\pi}\bigg[1-\mathrm
  e^{-\Gamma\tau_c}+f_1(\tau_c)\sum_{j>0}\frac{\Gamma_j}{\Gamma}\big(\cos(j\mathbf
  k\cdot\mathbf d)+1\big)\bigg]\nonumber\\
  &~~~~~~\times\langle\hat J^+(\mathbf k)\hat
  J^-(\mathbf k)\rangle_0,
  \label{final_result}
\end{align}
where
\begin{align}
  f_1(\tau_c)=&~(1-\mathrm
  e^{-\Gamma\tau_c})^2\frac{\alpha^{(0)}_0}{p^{(0)}_0}+\tau_c\mathrm
  e^{-\Gamma\tau_c}+\frac{1}{2}\mathrm e^{-2\Gamma\tau_c}-\frac{1}{2},\nonumber\\
  f_2(\tau_c)=&~\frac{1}{2}(1-\mathrm
  e^{-\Gamma\tau_c})^2\big\langle(\hat \sigma_m^z+1)(\hat
  \sigma_n^z+1)\big\rangle_0\nonumber\\
  &+(1-\Gamma\tau_c\mathrm e^{-\Gamma\tau_c}-\mathrm
  e^{-\Gamma\tau_c})(p^{(0)}_0-a^{(0)}_0)\nonumber\\
  &-(1-\mathrm
  e^{-\Gamma\tau_c})^2\Big(\frac{\alpha^{(0)}_0}{p^{(0)}_0}+\frac{1}{2}\Big)
  \big(a^{(0)}_0+p^{(0)}_0\big),\nonumber\\
  f_3(\tau_c)=&-(1-\mathrm
  e^{-\Gamma\tau_c})^2\Big(\frac{\alpha^{(0)}_0}{p^{(0)}_0}
  +\frac{1}{2}\Big)\big(a^{(0)}_0+p^{(0)}_0\big)\nonumber\\
  &+(1-\Gamma\tau_c\mathrm e^{-\Gamma\tau_c}-\mathrm
  e^{-\Gamma\tau_c})\big(a^{(0)}_0-p^{(0)}_0\big).
\end{align}
$f_1(\tau_c),f_2(\tau_c),f_3(\tau_c)\sim O\big((\Gamma\tau_c)^2\big)$
for $\Gamma\tau_c\ll1$.

Comparing Eq.~(\ref{0th}) and Eq.~(\ref{final_result}), we can see
that the modulation of the diffraction pattern by the multiple-light
scattering is described by $\cos(j\mathbf k\cdot\mathbf d)$. The
dipole-dipole interaction with coefficients $G_j$ has a vanishing
$1$-st order effect, thus it does not appear in our above
derivation. Obviously for $j\ll N$ the modulation is slowly varying in
$\mathbf k$ space. We are interested only in the diffraction pattern
in the neighborhood of the forward direction where $\cos(j\mathbf
k\cdot\mathbf d)\approx1$, then $n_p^{(1)}(\tau_c\to\infty)=\beta_1
n_p^{(0)}(\tau_c\to\infty)+\frac{\delta\Omega}{4\pi}N\beta_2$, with
$\beta_1=1+2f_1(\tau_c\to\infty)\sum_{j>0}\frac{\Gamma_j}{\Gamma}$ and
$\beta_2=[f_2(\tau_c\to\infty)+f_3(\tau_c\to\infty)]\sum_{j>0}\frac{\Gamma_j}{\Gamma}$. The
peak/dip to background ratio of the initial diffraction pattern
(i.e. $\tau_c\to0$) is
$r^{(0)}=\frac{N^2p^{(0)}_0}{N[a^{(0)}_0-p^{(0)}_0]}$ which measures
the pair-correlation sum of the initial atomic state of interest. In
the diffraction pattern of all emitted photons
(i.e. $\tau_c\to\infty$), it becomes
\begin{eqnarray}
  r^{(1)}&=&\frac{\beta_1N^2p^{(0)}_0+N\beta_2}{\beta_1N[a^{(0)}_0-p^{(0)}_0]+N\beta_2} 
  = (1+\delta)r^{(0)} ,\nonumber\\
  \delta &\cong &\frac{\beta_2}{\beta_1[a^{(0)}_0-p^{(0)}_0]+\beta_2}.
\end{eqnarray}
Since
$\sum_{j>0}\frac{\Gamma_j}{\Gamma}\approx\frac{1}{kd}\int_0^\infty\frac{\sin
  x}{x}dx\sim\frac{1}{kd}\ll1$, we expect $\beta_1\approx1$ and
$\beta_2\ll1$. Thus $\delta\ll1$.  When the initial atomic state is an
eigenstate of $\hat J_z$ or a separable state, we have
\begin{align}
  \delta \cong
  \frac{-2a^{(0)}_0p^{(0)}_0\sum_{j>0}\frac{\Gamma_j}{\Gamma}}{a^{(0)}_0-p^{(0)}_0}.
  \label{ratio_change}
\end{align}
We note that $p^{(0)}_0 = P/(N^2 - N)$, which has the maximum value of
$1/4$ in the neighborhood of Dicke state with
half-spin-excitation. For typical states, $p^{(0)}_0\sim 1/N$ and then
$\delta$ scales inversely with $N$.

\begin{figure}
  \includegraphics[bb=10bp 500bp 585bp 825bp,clip,width=8.5cm]{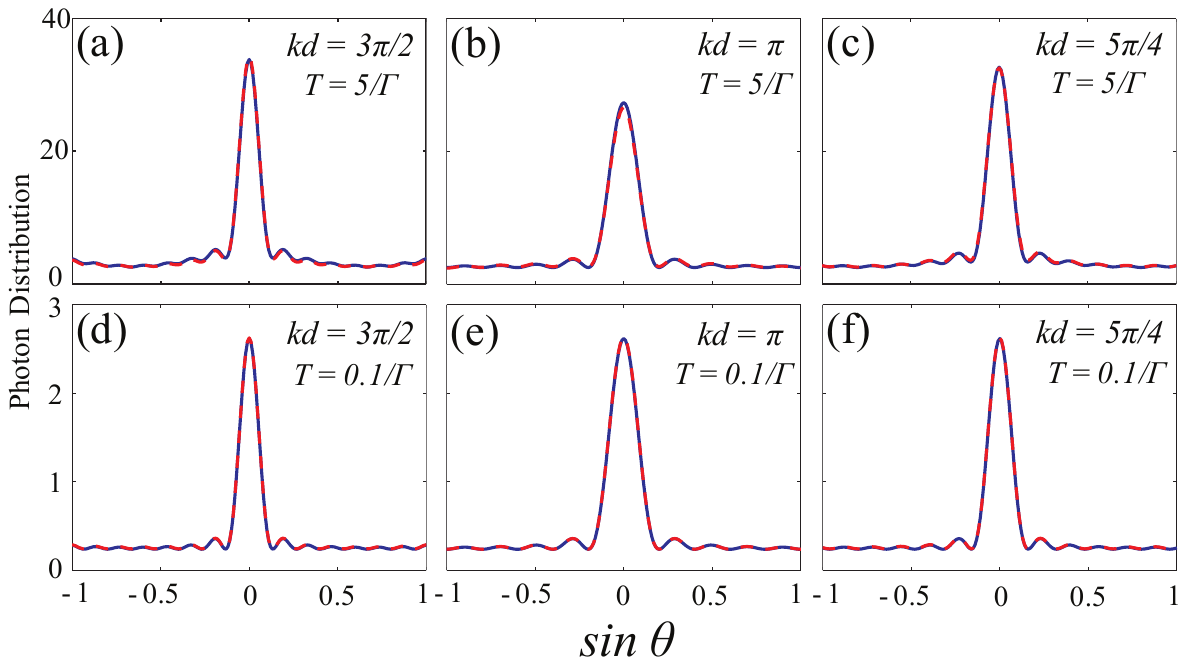}
  \caption{Diffraction pattern of Stokes photons emitted by a chain of
    $10$ atoms initially in the spin-coherent-state with in-plane
    polarization
    $\otimes_{j=1}^{10}\frac{|\uparrow\rangle_j+|\downarrow\rangle_j}{\sqrt{2}}$.
    With the collection interval $T=5/\Gamma$, $99.3\%$ of all
    spin-excitations are converted into Stokes photons, and with
    $T=0.1/\Gamma$, $10\%$ of all spin-excitations are converted into
    Stokes photons. Red dashed lines: exact numerical solution of the
    master equation Eq.~(\ref{original_ME}). Blue solid lines in
    (a-c): perturbative solution keeping the 1st order effect of
    $\mathcal{L}_1$ (i.e. Eq. (\ref{final_result})). Blue solid lines
    in (d-f): the $0$-th order solution without $\mathcal{L}_1$
    (i.e. Eq.~(\ref{0th})).  }
  \label{f3}
\end{figure}

\begin{figure}
  \includegraphics[bb=15bp 500bp 585bp 825bp,clip,width=8.5cm]{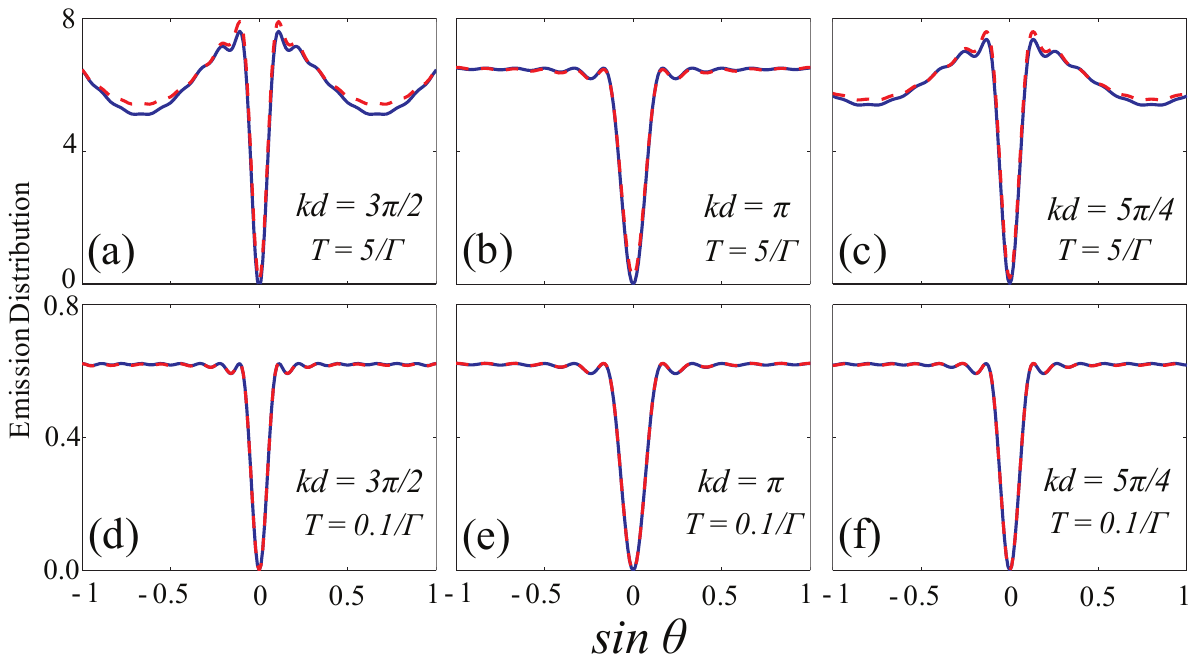}
  \caption{ Diffraction pattern of Stokes photons emitted by a chain
    of $12$ atoms initially in the many-body singlet (i.e. $J=0$,
    $M=0$). Red dashed lines: exact numerical solution of the master
    equation Eq.~(\ref{original_ME}). Blue solid lines in (a-c):
    perturbative solution keeping the 1st order effect of
    $\mathcal{L}_1$ (i.e. Eq. (\ref{final_result})). Blue solid lines
    in (d-f): the $0$-th order solution without $\mathcal{L}_1$
    (i.e. Eq.~(\ref{0th})).}
  \label{f4}
\end{figure}

\begin{figure}
  \includegraphics[bb=15bp 500bp 585bp 825bp,clip,width=8.5cm]{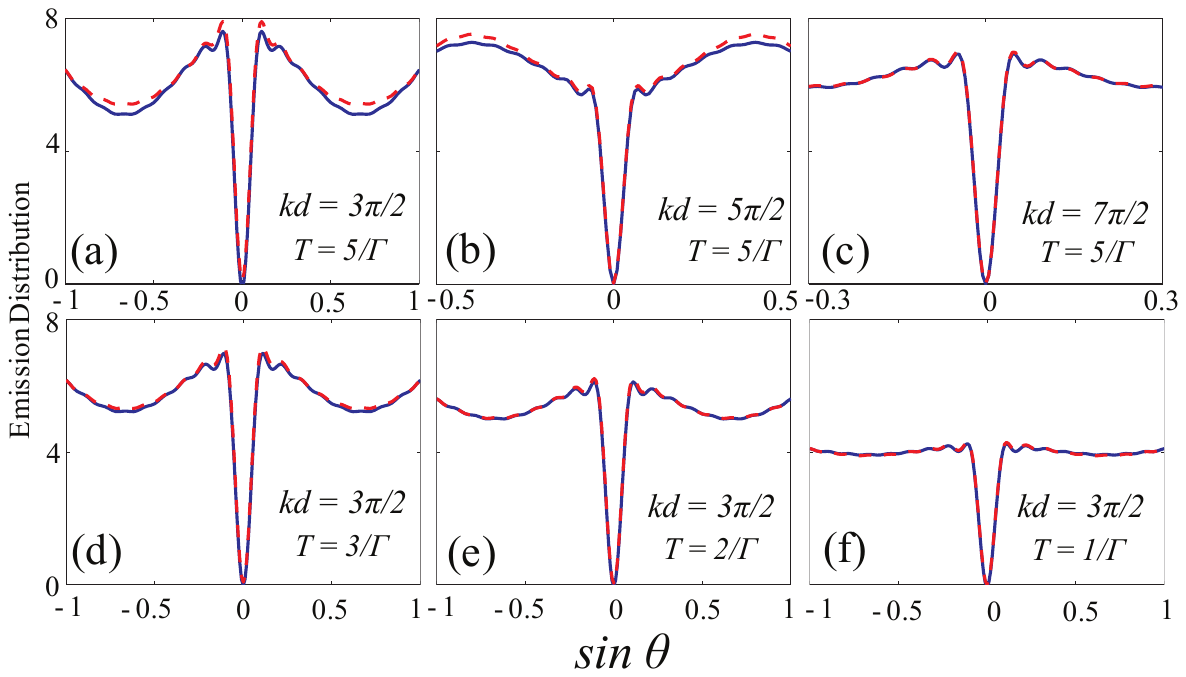}
  \caption{Diffraction pattern of Stokes photons emitted by a chain of
    $12$ atoms initially in the many-body singlet (i.e. $J=0$, $M=0$)
    with various duration of the collection time $T$. Red dashed
    lines: exact numerical solution of the master equation
    Eq.~(\ref{original_ME}). Blue solid lines: perturbative solution
    keeping the 1st order effect of $\mathcal{L}_1$
    (i.e. Eq. (\ref{final_result})). }
  \label{f5}
\end{figure}

\begin{figure}
  \includegraphics[bb=45bp 590bp 430bp 770bp,clip,width=8.5cm]{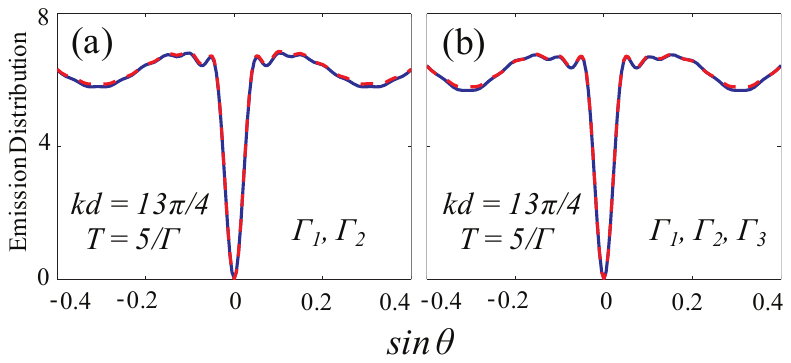}
  \caption{The results which not only includes the nearest neighbor
    $\Gamma_1$ term, but also contains (a) next-nearest neighbor
    $\Gamma_2$ and (b) $\Gamma_2$, $\Gamma_3$ terms for $12$-qubit
    $J=0$, $M=0$ permutation invariant state. Values of $d$ and $T$
    are given in the figure. Red dashed lines: the numerical simulated
    results. Blue solid lines: the fitting using
    Eq. (\ref{final_result}). The dipole-dipole interaction $G_j$ is
    not consider here.}
  \label{f6}
\end{figure}

To examine the convergence of the perturbation solution, we compare it
with exact numerical solution of the master equation for a small
ensemble in an 1D lattice with various lattice constant $d$. The
magnitude of the multiple-light scattering terms in $\mathcal{L}_1$
decays fast with the distance, thus only the nearest neighbor terms
are important. Because of the limit of computation capability, in the
calculation presented in Fig. \ref{f3}, \ref{f4} and \ref{f5} when
referring to multiple-light scattering, we only keep the nearest
neighbor terms in $\mathcal{L}_1$ (i.e. those with coefficients
$\Gamma_1$) and artificially set $\Gamma_j=0$ for $j\geq2$. But for
dipole-dipole interaction all $G_j$ terms are considered in the exact
numerical solution. For this reduced master equation, we compare the
perturbative solution and exact numerical solution. We can see the
perturbative solution keeping the 1st order effect of $\mathcal{L}_1$
(i.e. Eq. (\ref{final_result})) has excellent convergence to the exact
numerical solution for both the dip pattern and peak pattern. In
particular, the multiple light scattering has negligible effects if we
set the collection time $T \leq 0.1/\Gamma$.  Note that in the initial
interval of $T=0.1/\Gamma$, $10\%$ of all spin-excitations are already
converted into Stokes photons.  In Fig.~\ref{f6}, we show that effects
of next-nearest neighbor terms in $\mathcal{L}_1$ (i.e. with
coefficients $\Gamma_2$ and $\Gamma_3$) are also well accounted by the
perturbative solutions in Eq. (\ref{final_result}). This calculation
also confirms that the modulation of the diffraction pattern is
dominated by the near neighbor cross terms, and the effect of
$\Gamma_3$ term is already small as compared to the $\Gamma_1$ and
$\Gamma_2$ terms.

We have also considered a 2D atomic ensemble in the $xy$ plane in a
large trap, where the atomic number density is Gaussian $n(\mathbf
r)=n(0)\mathrm e^{-2r^2/A^2}$ at position $\mathbf r$ with $A$ the
full-width-half-maximum (FWHM). The total atom number is given by
$N=\int d\mathbf rn(\mathbf r)=\frac{\pi}{2}n(0)A^2$ with the dilute
condition $k_0 A \sqrt{N} \geq 2 \pi$ satisfied. The result is similar
to the case of 1D lattice that the peak/dip to background ratio is
only slightly changed.

For dilute hot atomic vapor where atomic motion is much faster than
the Stokes photon emission, we find that the diffraction pattern is
similar to that of cold atoms in the neighborhood of the forward
direction, i.e. with a sharp diffraction peak/dip from which the
pair-correlation sum of atoms can be read out. Since atoms move around
in a timescale faster than the photon emission, the coefficients of
cross terms in $\mathcal{L}_1$ should be replaced by
$\Gamma_{jj'}=\Gamma\big\langle\frac{\sin(k_0|\mathbf r_{j}-\mathbf
  r_{j'}|)}{k_0|\mathbf r_{j}-\mathbf
  r_{j'}|}\big\rangle_{\textrm{mot}}=\frac{\Gamma}{2(k_0A)^2}$,
$G_{jj'}=\frac{G}{(k_0A)^2}$ which are independent of $j$ and
$j'$~\cite{Wstate}. Here $A$ is the size of atomic vapor. The original
master equation Eq.~(\ref{original_ME}) becomes
\begin{align}
  \dot\rho=&~\frac{\Gamma}{2}\sum_{j}\left(2\hat\sigma_j^-\rho\hat
    \sigma_j^+-\hat\sigma_j^+\hat\sigma_j^-\rho-\rho\hat\sigma_j^+\hat\sigma_j^-\right)
  \nonumber\\
  &+\frac{\Gamma}{4(k_0A)^2}\left(2\hat J^-\rho\hat J^+-\hat J^+\hat
    J^-\rho-\rho\hat J^+\hat J^-\right) \nonumber\\
  &+i\frac{G}{2(k_0A)^2}[\hat J_x^2+\hat J_y^2,\rho].
\label{vapor_me}
\end{align}
In the RHS of above master equation, the first term corresponds to
atoms independently emit photons, the second term comes from multiple
light scattering, and the third term is the dipole-dipole
interaction. Then
\begin{align}
  \frac{d\langle\hat J_z\rangle_t}{dt}=-\Gamma\left(\frac{N}{2}+\langle\hat
  J_z\rangle_t\right)-\frac{\Gamma}{2(k_0A)^2}\langle\hat J^+\hat J^-\rangle_t,\nonumber\\
  \frac{d\langle\hat J^+\hat J^-\rangle_t}{dt}=-\Gamma\langle\hat
  J^+\hat J^-\rangle_t+\frac{\Gamma}{(k_0A)^2}\langle\hat J^+\hat J_z\hat
  J^-\rangle_t.
  \label{diff_eq}
\end{align}
The angular distribution of the emission rate is given by
\begin{align}
  &\langle\hat J^+(\mathbf k)\hat J^-(\mathbf
  k)\rangle_t\nonumber\\
  =&~\frac{N}{2}+\langle\hat J_z\rangle_t+\big(\langle\hat J^+\hat
  J^-\rangle_t-\langle\hat
  J_z\rangle_t-\frac{N}{2}\big)\big|\langle\mathrm e^{i\Delta\mathbf
    k\cdot\mathbf r_j}\rangle_\textrm{mot}\big|^2.
\end{align}

Just like the case of dilute cold atom ensemble, the initial
diffraction pattern has a sharp diffraction peak/dip in the forward
direction with a width given by $\frac{1}{k_0 A}$, and a strength
determined by the pair-correlation sum $P=\langle\hat J^+\hat
J^-\rangle_0-\langle\hat J_z\rangle_0-\frac{N}{2}$. The value of the
pair-correlation sum can be read out from the relative ratio of the
peak/dip to the neighboring background. From Eq.~(\ref{diff_eq}), we
can see the background part $\frac{N}{2}+\langle\hat J_z\rangle_t$ in
the emission rate decays with time, and the existence of
$\frac{\Gamma}{2(k_0A)^2}\langle\hat J^+\hat J^-\rangle_t$ term makes
the decay faster. On the other hand, the peak/dip strength
$\langle\hat J^+\hat J^-\rangle_t-\langle\hat J_z\rangle_t-\frac{N}{2}$
may increase with time as long as $\frac{\langle\hat J^+\hat J_z\hat
  J^-\rangle_t}{(k_0A)^2}+\frac{\langle\hat J^+\hat
  J^-\rangle_t}{2(k_0A)^2}>\langle\hat J^+\hat J^-\rangle_t-\langle\hat
J_z\rangle_t-\frac{N}{2}$. This is just the case discussed in the study
of superradiance phenomena by Rehler and
Eberly~\cite{Eberly_superradiance}, where the authors show that in a
very dense ensemble a directional superradiance may develop at later
time even when the initial state emission pattern is almost isotropic
in all direction and has no superradiance behavior.

Here we are interested in how the peak/dip to background ratio evolves
as a function of collection interval. It is obvious that only the
second terms in the RHS of both equations in Eq.~(\ref{diff_eq}) can
change this ratio. When the dilute condition $(k_0A)^2 / N\geq (2
\pi)^2$ is satisfied, $\frac{\langle\hat J^+\hat
  J^-\rangle_t}{2(k_0A)^2}\ll\frac{N}{2}+\langle\hat J_z\rangle_t$ and
$\frac{\langle\hat J^+\hat J_z\hat
  J^-\rangle_t}{(k_0A)^2}\ll\langle\hat J^+\hat J^-\rangle_t$, the
effects is negligible as compared to the first terms. Thus, the
peak/dip to background ratio is barely changed by the multiple light
scattering and dipole-dipole interaction when the dilute condition is
satisfied.

\section{Photon number fluctuations}
\label{fluctuation}

We analyze the shot noise of the photon counts at the detectors. The
total number of Stokes photons in a given direction $\mathbf k$ at
collection time $\tau_c$ is given by $n_p(\tau_c)=\sum_{\mathbf
  k}\langle\hat a^\dag_{\mathbf k}(\tau_c)\hat a_{\mathbf
  k}(\tau_c)\rangle$, here the summation is over a finite solid angle
$\delta\Omega$. From the effective light-atom coupling Hamiltonian
(Eq.~(\ref{light_atom_Hamiltonian})), the evolution of the photon
operator writes
\begin{eqnarray}
  \hat a_{\mathbf k}(t)=\hat a_{\mathbf k}(0)\mathrm
  e^{-i\omega_kt}\!+ig_{\mathbf k}\!\!\int_0^t\!\!d\tau\hat
  J^-(\mathbf k,\tau)\mathrm e^{-i\omega_{\mathbf k}(t-\tau)}.
    \label{akt}
\end{eqnarray}

The photon number fluctuation can then be expressed in terms of the
atomic correlations
\begin{eqnarray}
  \Delta n_p^2&\equiv&\Big\langle\Big(\sum_{\mathbf k}\hat
  a^\dag_{\mathbf k}(\tau_c)\hat a_{\mathbf
    k}(\tau_c)\Big)^2\Big\rangle-\Big\langle\sum_{\mathbf k}\hat
  a^\dag_{\mathbf k}(\tau_c)\hat a_{\mathbf k}(\tau_c)\Big\rangle^2\nonumber\\
  &=&n_p-n_p^2+\sum_{\mathbf k\mathbf k'}\big\langle\hat a^\dag_{\mathbf
    k}(\tau_c)\hat a^\dag_{\mathbf k'}(\tau_c)\hat a_{\mathbf k}(\tau_c)\hat
  a_{\mathbf k'}(\tau_c)\big\rangle\nonumber\\
  &=&n_p-n_p^2+\frac{2}{(4\pi)^2}\int\! d\Omega_{\mathbf
    k}d\Omega_{\mathbf
    k'}\int_0^{\tau_c}\!\!dt_1\!\int_0^{t_1}\!\!\!dt_2\nonumber\\
  &&\big\langle\hat{J}^+(\mathbf
  k',t_2)\hat{J}^+(\mathbf k,t_1)\hat{J}^-(\mathbf
  k,t_1)\hat{J}^-(\mathbf k',t_2)\big\rangle.
  \label{photonfluc}
\end{eqnarray}
Here for simplicity we have ignored the slowly varying single atom
dipole emission pattern.

In Appendix~\ref{perturbation}, we have shown that the $\mathcal{L}_1$ term in the
master equation only results in a small modulation on the diffraction
pattern. Thus, in deriving $\Delta n_p^2$ the effect of
$\mathcal{L}_1$ term will not be considered since we are only
interested in the order of magnitude of the fluctuation. From the
quantum regression theorem \cite{Carmichael_book}, we have
$\langle\hat{J}^+(\mathbf k',t)\hat{J}^+(\mathbf
k,t+\tau)\hat{J}^-(\mathbf k,t+\tau)\hat{J}^-(\mathbf k',t)\rangle
=\mathrm e^{-\Gamma\tau-2\Gamma t}\textrm{Tr}[\hat J^+(\mathbf k')\hat J^+(\mathbf
k)\hat J^-(\mathbf k)\hat J^-(\mathbf k')\rho(0)]$. Assuming that a
CCD pixel collects photons emitted in a small solid angle
$\delta\Omega\lesssim\frac{1}{N}$. During the time interval of $0$ to
$\infty$, we find the expectation value and fluctuation in the number
of photons collected by a pixel placed in the direction of $\mathbf
k$:
\begin{eqnarray}
  n_p&\approx&\frac{\delta\Omega}{4\pi}\textrm{Tr}\big[\hat{J}^+(\mathbf k)
  \hat{J}^-(\mathbf k)\rho(0)\big],\nonumber\\
  \Delta n_p^2&\approx&\frac{\delta\Omega^2}{(4\pi)^2}\textrm{Tr}\big[\hat J^+(\mathbf
  k)\hat J^+(\mathbf k)\hat J^-(\mathbf k)\hat J^-(\mathbf
  k)\rho(0)\big]\nonumber\\
  &&+n_p-n_p^2.
\end{eqnarray}

For atomic ensemble initially in eigenstates of $\hat J_z$ or
spin-coherent-states with in-plane polarization, a straightforward
calculation shows $n_p\sim N^2\delta\Omega\sim N$, $\Delta n_p^2\sim
n_p$ for $\mathbf k$ in the neighborhood of the forward direction, and
$n_p\sim N\delta\Omega\sim1$, $\Delta n_p^2\sim n_p$ for $\mathbf k$
away from the forward direction, i. e., the photon statistics is
Poissonian.

\section{Extraction of the peak position from the photon statistics}
\label{sen_diffract}

The most intuitive way to define the central position from the photon
distribution $\{ n_i + \Delta n_i\}$ collected at the CCD pixels is,
\begin{align}
  \theta_c=\frac{\sum_i\theta_i(n_i+\Delta n_i)}{\sum_i(n_i+\Delta
    n_i)}=\bar\theta_c+\frac{\sum_i(\theta_i-\bar\theta_c)\Delta
    n_i}{\sum_i(n_i+\Delta n_i)}. \label{peakcenter}
\end{align}
Where $\theta_c$ is the peak central position extracted from a single
probe, and $\bar{\theta}_c\equiv\frac{\sum_i\theta_in_i}{\sum_in_i}$
is the expectation value of $\theta_c$ in an ensemble measurement
consists of many probes. We first show that $\bar{\theta}_c$ can
infinitely approach $\theta_0$ with sufficient resolution of the CCD.

For large $N$, we can neglect the homogeneous background which is by a
factor of $N$ smaller than the peak feature, and write
\begin{align}
  \bar\theta_c=\frac{\sum_i\theta_i\bar f(\theta_i)}{\sum_i\bar
    f(\theta_i)}=\frac{\int_{\theta_\textrm{min}}^{\theta_\textrm{max}}d\theta\theta
    f(\theta)}{\int_{\theta_\textrm{min}}^{\theta_\textrm{max}}d\theta
    f(\theta)}+\epsilon=\theta_0+\epsilon.
\end{align}
The deviation $\epsilon$ comes from transforming the summation into
integral, which writes
\begin{align}
  \epsilon=&~\frac{\sum_i\int_{\theta_i-\frac{\delta\theta}{2}}^{\theta_i+\frac{\delta\theta}{2}}d\theta(\theta_i-\theta)f(\theta)}{\int_{\theta_\textrm{min}}^{\theta_\textrm{max}}
    d\theta f(\theta)} \nonumber\\
  =&~\frac{\sum_i\int_0^{\delta\theta/2}d\theta\theta\big[f(\theta_i+\theta)
    -f(\theta_i-\theta)\big]}{\int_{\theta_\textrm{min}}^{\theta_\textrm{max}}d\theta
    f(\theta)}\nonumber\\
  \approx&\frac{2\int_0^{\delta\theta/2}d\theta\theta^2\sum_i\frac{df(\theta)}{d\theta}\big|_{\theta=\theta_i}}
  {\int_{\theta_\textrm{min}}^{\theta_\textrm{max}}d\theta
    f(\theta)}. \label{correction}
\end{align}
Because function $f(\theta)$ satisfies
$\frac{df(\theta_0+\theta)}{d\theta}=-\frac{df(\theta_0-\theta)}{d\theta}$,
for every pixel index $i$ we can find $j$ such that
$\theta_i-\theta_0=\theta_0-\theta_j+O(\delta\theta)$, so
$\frac{df(\theta)}{d\theta}\big|_{\theta=\theta_i}+\frac{df(\theta)}{d\theta}\big|_{\theta=\theta_j}\sim(k_0A)^2f(\theta_i)\delta\theta+O(\delta\theta^2)$.
Thus the deviation in Eq.~(\ref{correction}) becomes, 
\begin{align}
  \epsilon\sim(k_0A)^2\int_0^{\delta\theta/2}d\theta\theta^2\sim(k_0A)^2\delta\theta^3,
\end{align}
which is $O(\delta\theta^3)$. When $\delta \theta \geq \frac{1}{k_0
  A}$, the sensitivity is determined by the CCD resolution. When
$\delta \theta \ll \frac{1}{k_0 A}$, $\bar{\theta}_c$ can infinitely
approach $\theta_0$.

In a single probe, the sensitivity is limited by the photon shot noise
which leads to uncertainty of $\theta_c$:
\begin{align}
\sqrt{ (\theta_c - \bar{\theta}_c)^2}\approx&~\frac{\sqrt{\sum_i(\theta_i-\bar{\theta}_c)^2 \Delta
      n^2_i}}{\sum_in_i}\nonumber\\
  \approx&~\frac{2}{N\sqrt{k_0A}}\frac{\sqrt{\int
      dxx^2\mathrm e^{-x^2/4}}}{\int dx\mathrm
    e^{-x^2/4}}\nonumber\\
  =&~\frac{2 \pi^{-1/4}}{N\sqrt{k_0A}}.
\end{align}
Thus, for using Eq.~(\ref{peakcenter}) to extract the Zeeman field
gradient $\partial_x \eta$ from a single probe, the overall precision
is
\begin{equation} 
  \Delta (\partial_x \eta) \sim \frac{k_0}{\tau_0} \sqrt{\frac{4\pi^{-1/2}}{N^2k_0A}+(k_0A)^4\delta\theta^6}.
\end{equation}
For small $\delta\theta$, the sensitivity is $\Delta (\partial_x \eta)
\sim \frac{k_0}{\tau_0} \frac{1}{N\sqrt{k_0A}}$ which
scales inversely with $N$.

We can also use a function $g(\theta_i,\alpha)=\delta\theta
\frac{N^2}{4}\mathrm
e^{-\frac{(k_0A)^2}{4}(\theta_i-\alpha)^2}+\delta\theta \frac{N}{4}$
to fit the obtained data $n_i+\Delta n_i$. The peak position
$\theta_c$ is defined as the value of $\alpha$ which minimizes
$\sum_i\big[g(\theta_i,\alpha)-n_i-\Delta n_i\big]^2$. This method
gives the same sensitivity $\Delta (\partial_x \eta) \sim
\frac{k_0}{\tau_0} \frac{1}{N\sqrt{k_0A}}$ for small $\delta\theta$.

\section{Gradiometer Using Flying Atom Mach-Zehnder Interferometry}
\label{sen_MZI}

\begin{figure}[t]
  \includegraphics[width=5cm]{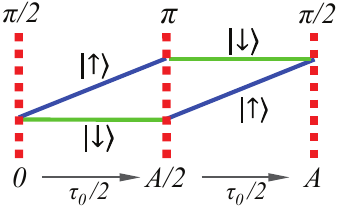}
  \caption{Gradiometer based on flying atom Mach-Zehnder interferometry (MZI).}
\label{f10}
\end{figure}

 A standard method for field gradiometer is based on phase estimation
 using flying atom Mach-Zehnder interferometer
 (MZI)~\cite{MZI_gradiometer1} (see Fig.~\ref{f10}). First, a $\pi/2$
 pulse is applied to prepare an atom in the superposition of spin-up
 and spin-down state, which is then launched from $x=0$ with a velocity
 $v$.  After free evolution for an interval $\tau_0/2$ , a $\pi$ pulse
 is applied to flip the spin. Finally, another $\pi/2$ pulse is applied
 and the population on the spin-up state is measured for observing the
 interference signal. In the first interval of $\tau_0/2$, the spin up
 and down states acquire a relative phase shift
 $\phi_1=(\eta_0+\frac{v\tau_0\partial_x\eta}{4})\frac{\tau_0}{2}$,
 where $\eta_0$ is the homogeneous part of the Zeeman field and
 $\partial_x\eta$ is the gradient to be measured. $v \tau_0/2 = A/2$ is
 the distance travelled by the atom with velocity $v$. In the second
 interval of $\tau_0/2$, the Zeeman field induces a phase shift
 $\phi_2=-(\eta_0+\frac{3v\tau_0\partial_x\eta}{4})\frac{\tau_0}{2}$. By
 measuring the population of the spin up state, this MZI gives an
 estimate of the total phase
 $\phi\equiv\phi_1+\phi_2=-\frac{1}{4}v\tau_0^2\partial_x\eta$ from
 which the gradient is then inferred:
 $\partial_x\eta=-\frac{4\phi}{v\tau_0^2}$.

The sensitivity of the gradiometer is limited by the shot noise
$\Delta\phi$ for phase estimation, and the uncertainty in velocity
$\Delta v$. In the $i$-th probe using a single atom, the phase one
readout can be written as
\begin{align}
  \phi_i=\frac{1}{4}\partial_x\eta\tau_0^2(\bar v+\Delta v_i)+\Delta\phi_i.
\end{align}
where $\Delta\phi_i$ is the error for phase estimation. $\bar v$ and
$\Delta v_i$ are respectively the expectation value and uncertainty in
velocity. The inferred value of the gradient is then,
\begin{align}
  \partial_x\eta^{(i)}\equiv\frac{4\phi_i}{\tau_0^2\bar
    v}=\partial_x\eta+\partial_x\eta\frac{\Delta v_i}{\bar
    v}+\frac{4\Delta\phi_i}{\tau_0^2\bar v}.
\end{align}
Using an ensemble of $N$ atoms, the error in measuring the gradient is
\begin{align}
&\frac{1}{N} \sqrt{\sum_i (\partial_x\eta^{(i)} - \partial_x\eta)^2}\nonumber\\
=&~\frac{\partial_x\eta}{\bar
    v} \frac{1}{\sqrt{N}}  \sqrt{ \frac{1}{N} \sum_i\Delta v_i^2} + \frac{4}{\tau_0^2\bar v}  \frac{1}{\sqrt{N}} \sqrt{\frac{1}{N} \sum_i\Delta\phi_i^2}.
\end{align}
$\sqrt{\frac{1}{N} \sum_i\Delta\phi_i^2} = 1$ which is the shot noise
for phase estimation using the MZI~\cite{Lloyd_metrology}. Thus the
precision of MZI scheme is:
\begin{equation}
  \Delta(\partial_x\eta)=\partial_x\eta \frac{\Delta v}{\sqrt{N}\bar
    v} +\frac{4}{\sqrt{N}\tau_0 A},
  \label{MZIgradiometer}
\end{equation}
where $A = \bar{v} \tau_0 $ represents the spatial resolution of the
gradiometer and $\Delta v = \sqrt{ \frac{1}{N} \sum_i\Delta v_i^2} $
is the velocity uncertainty.  

For atoms launched by an atomic fountain, the uncertainty in atom
velocity is intrinsically limited by the temperature: $ \Delta v \sim
\sqrt{k_BT/m}$. For example, with a temperature of $T\sim1~\mu$K,
$\Delta v\sim1$ cm/s. To reduce the error caused by this velocity
uncertainty (1st term on RHS of Eq.~(\ref{MZIgradiometer})), $\bar{v}$
shall be large as compared to $\Delta v$. This then sets upper bound
for the probe time at a desired spatial resolution since $\tau_0 =
A/\bar{v}$.


\begin{thebibliography}{99}

\bibitem{Bloch_RMP}
  I. Bloch, J. Dalibard, and W. Zwerger,
  Rev. Mod. Phys. $\bf 80$, 885 (2008).

\bibitem{Guhne_entanglementdetection}
  O. Guhne, and G. Toth,
  Phys. Rep. $\bf 474$, 1 (2009).

\bibitem{Lloyd_metrology}
V. Giovannetti, S. Lloyd, and L. Maccone, Science $\bf306$, 1330 (2004).

\bibitem{EIT}
M. Fleischhauer, A. Imamoglu, and J. P. Marangos, Rev. Mod. Phys. $\bf 77$, 633, (2005).

\bibitem{DLCZ}
L.-M. Duan, M. D. Lukin, J. I. Cirac and P. Zoller, Nature $\bf 414$, 413 (2001).

\bibitem{quantum_memory0}
C. H. van der Wal \textit{et al.}, Science $\bf 301$, 196 (2003).

\bibitem{quantum_memory1}
B. Julsgaard, J. Sherson, J. I. Cirac, J. Fiurasek, and E. S. Polzik, Nature $\bf 432$, 482 (2004).

\bibitem{quantum_memory2}
D. N. Matsukevich \textit{et al.}, Phys. Rev. Lett. $\bf 97$, 013601 (2006).

\bibitem{quantum_memory3}
J. Simon, H. Tanji, J. K. Thompson, and V. Vuletic, Phys. Rev. Lett. $\bf 98$, 183601 (2007).

\bibitem{quantum_memory4}
C.-W. Chou \textit{et al.}, Science $\bf 316$, 1316 (2007).

\bibitem{Pan_quantummemory}
B. Zhao \textit{et al.}, Nat. Phys. $\bf 5$, 95 (2009).

\bibitem{dephaseinOL1}
  U. Schnorrberger \textit{et al.}, Phys. Rev. Lett. $\bf103$, 033003 (2009).

\bibitem{dephaseinOL2}
R. Zhao \textit{et al.}, Nat. Phys. $\bf5$, 100 (2009).

\bibitem{Eberly_superradiance}
N. E. Rehler and J. H. Eberly, Phys. Rev. A $\bf 3$, 1735 (1971).

\bibitem{Haroche_superradiance}
  M. Gross and S. Haroche, Phys. Rep. $\bf93$, 301 (1982).

\bibitem{Carmichael_superradiance}
  J. P. Clemens, L. Horvath, B. C. Sanders, and H. J. Carmichael, Phys. Rev. A
  $\bf68$, 023809 (2003).

\bibitem{Scully_singlephotonsuperradiance1}
M. O. Scully, E. S. Fry, C. H. Raymond Ooi, and K. Wodkiewicz, Phys. Rev. Lett. 96, 010501 (2006)

\bibitem{Scully_singlephotonsuperradiance2}
M. O. Scully and A. A. Svidzinsky, Science $\bf 325$, 1510 (2009).

\bibitem{Eberly_singlephotonsuperradiance}
J. H. Eberly, J. Phys. B: At. Mol. Opt. Phys. $\bf 39$, S599 (2006).

\bibitem{Wstate}
  D. Porras and J. I. Cirac, Phys. Rev. A $\bf 78$, 053816 (2008).

\bibitem{Wstate2}
  R. Wiegner, J. von Zanthier and G. S. Agarwal, Phys. Rev. A $\bf 84$, 023805 (2011).
  
\bibitem{Altman}
E. Altman, E. Demler, and M. D. Lukin, Phys. Rev. A $\bf 70$, 013603 (2004).

\bibitem{Eckert}
K. Eckert \textit{et al.}, Nature Physics $\bf 4$, 50 (2008).

\bibitem{Bruun}
G. M. Bruun, B. M. Andersen, E. Demler, and A. S. Sorensen, Phys. Rev. Lett. $\bf 102$, 030401 (2009).

\bibitem{Vega}
I. de Vega, J. I. Cirac and D. Porras, Phys. Rev. A $\bf 77$, 051804(R) (2008).

\bibitem{Corcovilos}
T. A. Corcovilos, S. K. Baur, J. M. Hitchcock, E. J. Mueller, and R. G. Hulet, Phys. Rev. A $\bf 81$, 013415 (2010).

\bibitem{Miyake}
H. Miyake \textit{et al.}, Phys. Rev. Lett. $\bf 107$, 175302 (2011).

\bibitem{Weitenberg}
C. Weitenberg \textit{et al.}, Phys. Rev. Lett. $\bf 106$, 215301 (2011).

\bibitem{Sorensen_spinsqueezing}
A. S. Sorensen and K. Molmer, Phys. Rev. Lett. $\bf 86$, 4431 (2001).

\bibitem{Korbicz_spinsqueezing}
J. K. Korbicz \textit{et al.}, Phys. Rev. A $\bf 74$, 052319 (2006).

\bibitem{Toth_squeezing}
  G. Toth, C. Knapp, O. Guhne and H. J. Briegel,
  Phys. Rev.  Lett. $\bf 99$, 250405 (2007).

\bibitem{Duan2011}
  L.-M. Duan, Phys. Rev. Lett. $\bf 107$, 180502 (2011).

\bibitem{Delocalized_entanglement}
  K. G. H. Vollbrecht and J. I. Cirac, Phys. Rev. Lett. $\bf 98$, 190502 (2007).

\bibitem{fockstates1} M. J. Holland and K. Burnett, Phys. Rev. Lett. $\bf 71$, 1355 (1993).

\bibitem{fockstates2} J. Jacobson, G. Bjork, I. Chuang, and Y. Yamamoto, Phys. Rev. Lett. $\bf 74$, 4835 (1995).

\bibitem{HL1}
B. L. Higgins, D. W. Berry, S. D. Bartlett, H. M. Wiseman and G. J. Pryde, Nature $\bf 450$, 393 (2007).

\bibitem{HL2}
D. Braun and J. Martin, Nat. Commun. $\bf 2$, 223 (2011).

\bibitem{Carmichael_book}
  H. J. Carmichael, \textit{An Open Systems Approach to Quantum
    Optics}, Lecture Notes in Physics, New Series: Monographs,
  Vol. m18 (Springer, Berlin, 1993).

\bibitem{MZI_gradiometer1}
M.-K. Zhou \textit{et al.}, Phys. Rev. A $\bf 82$, 061602(R) (2010).

\bibitem{MZI_gradiometer2}
J. B. Fixler, G. T. Foster, J. M. McGuirk and M. A. Kasevich, Science $\bf 315$, 74 (2007).

\end{thebibliography}
\end{document}